\documentclass[]{article}
\usepackage{graphicx}
\usepackage[hmargin=3cm,vmargin=3cm,bindingoffset=0.5cm]{geometry}
\usepackage{authblk}

\title{Detection of Spatiotemporally Coherent Rainfall Anomalies Using Markov Random Fields}
\author[1]{Adway Mitra}
\author[2]{Ashwin K. Seshadri}
\affil[1]{International Center for Theoretical Sciences (ICTS), Bangalore, India}
\affil[2]{Divecha Centre for Climate Change, IISc, Bangalore, India}

\begin{document}

\maketitle

\begin{abstract}
Precipitation is a large-scale, spatio-temporally heterogeneous phenomenon, with frequent anomalies exhibiting unusually high or low values. We use Markov Random Fields (MRFs) to detect extended anomalies in gridded annual rainfall data across India from 1901-2005, that are spatio-temporally coherent but permitting flexibility in size. MRFs are undirected graphical models where each node is associated with a \{location,year\} pair, with edges connecting nodes representing adjacent locations or years. Some nodes represent observations of precipitation, while the rest represent unobserved (\emph{latent}) states that can take one of three values: high/low/normal. The MRF represents a probability distribution over the variables, using \emph{node potential} and \emph{edge potential} functions defined on nodes and edges of the graph. Optimal values of latent state variables are estimated by maximizing their posterior probability using Gibbs sampling, conditioned on the observations. These latent states are used to identify spatio-temporally extended rainfall anomalies, both positive and negative. Edge potentials enforce spatial and temporal coherence, and can adjust the competing influences of these types of coherence. We study spatio-temporal properties of rainfall anomalies discovered by this method, using suitable measures. We also study the relations between spatio-temporal sizes and intensities of anomalies. Identification of such rainfall anomalies can help in monitoring and studying floods and droughts in India. Additionally, properties of anomalies learnt from this approach could present tests of regional-scale rainfall simulations by climate models and statistical simulators. 
\end{abstract}

\section{Introduction} 
In many parts of the world, such as India, rainfall plays an important role in the economy and the well-being of millions of people. Consequently, excess or deficient rainfall can have very significant effects, especially if it is spread over a large region, or a long time. It is known that low annual rainfall has an adverse effect on India's GDP \cite{gdp}. Hence, identification of such spatio-temporally extended events of excess or deficient rainfall is important in both observed historical data and simulations of future scenarios by climate models. In this work, we call such events ``anomalies".

In climate science, ``anomaly" of a climatic variable (such as precipitation) at a particular location and time is defined quantitatively, as the amount of deviation from its climatological value, averaged over many years. But in this work, we will use the term ``anomaly" to indicate deviation not only from climatological value at individual locations but also with respect to spatial/temporal neighbors. Instead of individual locations such as grid-points in individual years, we consider spatially or temporally extended anomalies, with flexible spatio-temporal sizes. Anomalies can occur at different spatial and temporal scales, and their occurrence is heterogeneous (the statistics are location-dependent) and anisotropic (not uniform in all directions). The more consequential anomalies are the ones with significant spatiotemporal extent, and therefore it is important to identify them. Identification of such anomalies of rainfall are very useful in monitoring floods~\cite{flood} and droughts~\cite{drought1,drought2,drought3} in India, as it gives us the information about which regions received excess or deficient rainfall in any given year. Of course, floods and droughts can occur at sub-annual time scales, and any approch to detection of such anomalies should be general enough to work at any time scale of interest. Another factor is that with climate change, the frequency of rainfall extremes may increase, along with changes in the spatial pattern of rainfall\cite{spatvar}. To understand past and future changes, scientists rely on climate models like general circulation models (GCMs) which simulate global climatic variables including rainfall. Algorithms are necessary for analyzing large-scale simulations as well as observational data procured from sensors, and such analyses should include detecting and summarizing statistics of rainfall anomalies (\cite{abrupt,wavelet,ICA}). Such analysis cannot be done manually because of the large and growing volume of data and simulation results, raising the need for automated procedures.

Automating anomaly detection is challenging, because anomalies are inherently subjective, depending on definition and detection threshold \cite{anomalysurvey}. Anomaly detection in general, and  spatiotemporal anomaly detection in particular are considered important research areas in Data Science \cite{STDM}. Anomalies can be both positive and negative depending on the sign of deviation of rainfall volume from the long-term mean. However, the magnitude of deviation to be considered as ``anomaly" is a design choice. The simplest approach to anomaly detection is based on a predefined threshold, relative to statistics of the corresponding variable at individual spatial locations. With rainfall, one might consider the time-series of annual mean rainfall at each grid location, estimate its mean and variance, and identify years departing significantly from the mean. However, accounting for effects of spatiotemporal neighbours is important for detection \cite{droughtstudy} of extended anomalies, and the aforementioned location-wise threshold-based approach cannot do this. Neither is it suitable to establish fixed thresholds for spatio-temporal sizes defining anomalies, as these have a wide range of sizes. Several spatially separated anomalies are present within the same year, some of which may be of different signs. Basically, we need to make a compromise between the magnitude and spatio-temporal extent. 

Anomalies can occur at different spatial scales, ranging from that of the entire domain, in this case the country scale, down to grid levels. An anomaly of all-India rainfall is likely to be manifested through several smaller anomalies of the same sign. For example if the entire country has a negative anomaly in a given year,  then several grid-locations within the country are likely to be parts of negative anomalies during the same year. The Indian Meteorological Department (IMD) declares years to be ``excess rainfall" (positive anomaly), ``deficient rainfall" (negative anomaly) or normal, by comparing the aggregate all-India annual rainfall against thresholds. In some applications, whether or not an anomaly is identified at a large spatial scale should also depend on the presence/absence of anomalies at smaller scales, and the methods illustrated here facilitate this. For example, a year with widespread drought and many grid-locations under negative anomalies, could be considered as a year of negative anomaly at the all-India scale, even if all-India rainfall were not below the threshold.

Furthermore the anomaly detection problem is broader, especially when the anomaly is conceived as a conceptual or abstract quantity represented by a state variable that cannot be directly observed or measured and must be inferred indirectly. Here we consider anomalies in rainfall as a \emph{latent variable}, as often done in statistical modelling \cite{bishop} including spatiotemporal modelling \cite{spatstat}. Such latent (i.e. unobserved) states are best estimated through probabilistic methods \cite{bishop,neal}. We associate a latent state variable with each spatiotemporal location, i.e. each combination of grid-point and year. A graph is constructed with all these spatio-temporal variables as nodes, where pairs of nodes corresponding to neighboring locations are connected by edges. An anomaly is a connected component of such a graph, such that at each node in the component the associated latent variables have equal value. The approach of using local wet/dry conditions along with their spatio-temporal extents for monitoring floods and droughts has been attempted earlier also~\cite{local}, though using standard precipitation index instead of discrete variables.

In this work we model these latent variables to be spatiotemporally coherent through parameters of a Markov Random Field. We estimate these latent variables as the maximum posterior (MAP) solution of a Markov Random Field (MRF). MRFs are undirected random graphical models satisfying Markov properties, and are generally used to model joint distributions of several variables \cite{MRF}. Given a likelihood model of the data conditional on the states of this graph, the posterior density and correspondingly the MAP solution of these variables can be estimated.  Each latent state node has three values: 1 (positive anomaly), 2 (negative anomaly) or 3 (normal).  We also have additional nodes for all-India states each year, which are connected to the local nodes to account for the interaction between spatial scales. MRFs are defined using ``potential functions" for nodes and edges of the graph, which encode interactions between neighbouring variables. In our application, these functions influence the spatial and temporal coherence of the state variables. The local Markov properties inherent to MRFs imply that, for any node, its value is conditionally independent of all other nodes except its neighbours.

To identify the MAP configuration of the latent states  we use Gibbs sampling. Based on the inferred latent states we identify spatiotemporally coherent anomalies, and quantify their properties. Effects of enforcing spatial and temporal coherence on the resulting anomalies are examined, and sensitivity to parameters is studied. We compare the spatial extents of positive and negative anomalies. There is an inherent trade-off between spatial and temporal extents of anomalies in any procedure, originating in the values of parameters enforcing spatial and temporal coherence. Furthermore, even for any fixed set of parameters, there is variability in the spatial and temporal sizes of the anomalies detected across the spatio-temporal domain. Both of these effects are examined. Finally, we also study the intensity of anomalies, i.e.the degree by which the annual rainfall in a set of locations suffering an anomaly differs from the long-term rainfall there. We also study how this intensity is related to spatial and temporal extents of the anomalies. Somewhat similar properties of droughts have been studied earlier~\cite{droughtprop}, with an aim to filter out minor droughts. We illustrate our analysis with case studies of some spatially and temporally extended anomalies that our method detected.

The contribution of this paper is to study a new problem - detection of spatio-temporally extended rainfall anomalies. We cast the problem into the anomaly detection framework of Data Mining, and use probabilistic approach based on mixture models and latent variables. We use Markov Random Fields for spatio-temporal coherence. A major advantage of this approach is that no thresholds are needed, and anomalies of arbitrary shapes and sizes can be detected. Also, we consider the interaction between different spatial scales. The properties of the model are studied extensively.

\section{Methodology}
\subsection{Definitions and Notation}We consider $S$ locations and $T$ years, and spatiotemporal observations $Y_{st}$ of a geophysical variable such as annual-mean rainfall. Then $s$ indexes location and $t$ indexes time, and $Y_{st}$ signifies rainfall received by location $s$ at time $t$. Unlike time, 2-dimensional spatial locations have no natural ordering. So we order the spatial locations based on their longitude first, latitude next.  Each location in the 2-dimensional spatial grid system has 8 neighbors. For each location $s$, we denote by $NB(s)$ the set of its neighboring locations, according to the grid system. Thus, for a location having coordinates $(lat,lon)$, its neighbors will be $\{(lat+i,lon+j)\}$, where $i,j \in \{-1,0,1\}$. This particular way of ordering and indexing the spatial locations has no bearing on the analyses undertaken below, and any other indexing scheme is also equally compatible with it. This is because, the indexing does not indicate any sequence of the spatial locations, it just identifies them. The important thing in our analysis is the neighborhood structure, which is based on the spatial locations of the grids and independent of the indexing scheme.

Let us consider a graph $G$, where each node is associated with a pair $(s,t)$.  Further, for each spatio-temporal location $(s,t)$ we have two nodes, one corresponding to $Z_{st}$ and one for $Y_{st}$. $Z_{st}$ is a discrete variable which indicates the state of rainfall at location $s$, time $t$. While $Y$ is known from the dataset, $Z$ is unknown, and must be estimated. We put edges between pairs of nodes corresponding to $Z_{st}$ and $Z_{s't}$ for each year $t$ if $s$ and $s'$ are neighbouring grid-points, i.e. $s' \in NB(s)$. We call such edges as \emph{spatial edges}. Again, we put edges between pairs of nodes corresponding to $Z_{st}$ and $Z_{s,t+1}$ for each location $s$, and such edges are called \emph{temporal edges}. Finally, for each spatio-temporal pair $(s,t)$ we have an edge between $Z_{st}$ and $Y_{st}$, and we call such edges as \emph{data edges}. Thus a spatial edge connects a $Z$-nodes associated with neighboring locations and same time, a temporal edge connects $Z$-nodes associated with same location but adjacent times, and a data-edge connects $Z$-node and $Y$-node at the same location and time. Thus, we have $2ST$ nodes, $ST$ data edges, $S(T-1)$ temporal edges, and $\sum_{s}|NB(s)|T$ spatial edges.

We consider each location $s$ to be in one of three possible \emph{states} in any year $t$- high (1), low (2) or normal (3), which is encoded by $Z$. This follows the conventional classification of rainfall-years as excess rainfall, deficient rainfall, or normal, at each location. The state is represented by a latent discrete variable $Z_{st}$ taking one of 3 values. In such a graph, an \emph{anomaly} is a connected component of the $Z$-nodes corresponding to spatio-temporal locations,  such that all of the nodes in the component have the same value of $Z$ : either 1 (positive anomaly) or 2 (negative anomaly).  A goal of anomaly detection is to estimate these latent variables, from which the connected components can be computed and thus spatio-temporally coherent anomalies identified \cite{anomalysurvey}.

\subsection{Location-wise Analysis (LWA)}
A naive solution to anomaly detection is to treat the time-series at each location individually. For each time-series we compute mean $\mu_s$ and standard deviation $\sigma_s$. We then set $Z_{st}=1$ (high) for those years where $Y_{st} \geq HIGH_s$, $Z_{st}=2$ (low) for those years where $Y_{st} \leq LOW_s$, and $Z_{st}=3$ (normal) for all other years, where $HIGH_s$ and $LOW_s$ are thresholds specific to location $s$. We call this method Location-Wise Analysis (LWA), since it treats each location independently without considering the state of its neighbours. Corresponding assignments to the latent variables by this method are denoted as $Z0$. 

This approach suffers from two major limitations. Firstly, it is not clear how to choose the thresholds, and results vary strongly with the choice. The histogram of annual rainfall in most locations resembles the bell-shaped curve of Gaussian distribution. So, it is reasonable to set $HIGH_s=\mu_s+\sigma_s$ and $LOW_s=\mu_s-\sigma_s$. Through the rest of this paper, we will use this choice.  However, an approach that circumvents the need to specify such thresholds is a better solution.

The second major limitation of this approach is of course its neglect of spatial coherence in the latent variable. For example an individual location may be in a certain mode, while all its neighbours are in a different mode in the same year. Isolated anomalies need not be spurious, but spatially or temporally extended anomalies are more consequential.   An alternate approach might be to undertake location wise analysis, after having smoothed data onto a coarser grid. This enlarges the scales of interest, but involves loss of spatial information. It also does not permit anomalies at multiple scales, or naturally accommodate spatial heterogeneity or anisotropy in anomalies. This is the most important limitation of LWA, and we need a fundamentally new approach to circumvent it.

Finally, this approach also neglects temporal coherence in each of the location-specific time-series. This shortcoming can be solved by using an approach like Hidden Markov Models, which consider a discrete state space for a time-series and models the state transition distributions. However, Hidden Markov Models are most suitable when there exists some natural ordering between the states and one particular state is likely to be followed by another state. In this case, we do not have any such ordering. Rather, we simply need state persistence to achieve temporal coherence. This can be achieved by the method proposed below.

\subsection{Modeling by Markov Random Fields}
Detecting extended anomalies requires a different lens from LWA, one inducing spatial or temporal coherence during assignment of the $Z_{st}$-variables.  To address this shortcoming, we take the approach that assigns probabilities to different configurations of latent $Z$-variables, with higher weights to configurations where $Z$-assignments are spatially or temporally coherent. This is achieved by modelling the latent variable as an MRF, along the lines of the drought discovery technique in \cite{MRFdrought}. We seek to discover spatial and temporal clusters within which $Z$-values are the same.

Markov Random Field is an undirected graphical model, where a probability distribution are defined on an undirected graph. Each node in the graph corresponds to a random variable, and each edge has an associated potential function that depends on the random variables corresponding to the two nodes connected by that edge. The full likelihood of the model is defined as the product of all the edge potential functions.

As already stated, we have 2 nodes for every spatio-temporal pair $(s,t)$ - corresponding to $Z_{st}$ and $Y_{st}$. Spatial edges, temporal edges and data edges are defined between pairs of variables as mentioned above. In addition to grid-wise latent states, these can also be defined for the all-India mean, relative to its corresponding distribution across years. The Indian Meteorological Department (IMD) currently makes annual forecasts of spatial aggregate rainfall over India during the summer monsoon months of July-September (JJAS), called Indian Summer Monsoon Rainfall (ISMR). We define an analogous quantity for the entire year, All-India Mean Rainfall (AIMR), and denote by $Y_t$. Its anomalies are relative to its interannual mean $\mu$ and standard deviation $\sigma$. Once again, we define discrete latent variable $Z_t$ corresponding to AIMR, which can take 3 values.

A Markov Random Field is an undirected graph, with nodes for each $(s,t)$ pair. Corresponding to each $(s,t)$ pair is associated a latent variable $Z_{st}$ and an observation $Y_{st}$.  Each observation node $Y_{st}$ has a single edge, to the corresponding latent variable node $Z_{st}$. The graph also contains nodes corresponding to each year, associated with latent $Z_t$ and observed $Y_t$, corresponding to AIMR. For any year $t$, $Z_t$ is linked by edges to all nodes for that year $\{Z_{st}\}$ for every location $s$. Large anomalies in ISMR are declared by IMD as excess or deficient rainfall years. However, rainfall is highly heterogeneous spatially. Therefore in order to define anomalies in the aggregate measure of AIMR, we consider not only calculations of $Y_t$ but also the frequencies of local anomalies in the corresponding year. This is achieved by linking the $Z_{st}$ and $Z_t$ nodes. Figure 1 illustrates the model.

Probabilities are assigned to each configuration of $Z$ using node \emph{potential functions} $\psi^v(Z_{st})$ on each node, edge potentials  $\psi^e(Z_{st},Z_{s't'})$ on each edge occurring between spatiotemporal nodes and $\psi^f(Z_{st},Z_{t})$ on each edge occurring between spatiotemporal nodes and AIMR nodes. Edge potentials influence spatial and temporal coherence and node potentials influence the threshold for anomaly detection. Edge potentials describe prior probabilities that the nodes connected by the edge are in the same state. The node potential functions can be interpreted as describing the prior probability distribution across different states.

The precipitation amount at any location and year, given by $Y_{st}$, is modelled using a Gaussian distribution with parameters specific to the location $s$ and latent state $Z_{st}$. These conditional distributions can be interpreted as edge potentials on the $Z_{st}-Y_{st}$ data edges connecting the latent and observed states respectively.

The likelihood function is:
\begin{equation}L(Z) \propto \prod\limits_{s,t}{\psi^v(Z_{st})}\prod\limits_{e}{\psi^e(Z_{st},Z_{s't'})}\prod\limits_{f}{\psi^f(Z_{st},Z_{t})} \prod_{s,t}\mathcal{N}(Y_{st}; \mu_{sZ_{st}},\sigma_{s})\prod_{t}\mathcal{N}(Y_{t}; \mu_{Z_{t}},\sigma) \nonumber
\end{equation}
This defines the likelihood function, i.e. the probability of observing the data given the latent variables in the graph.

\begin{figure}
	\centering
	\includegraphics[width=12cm,height=6cm]{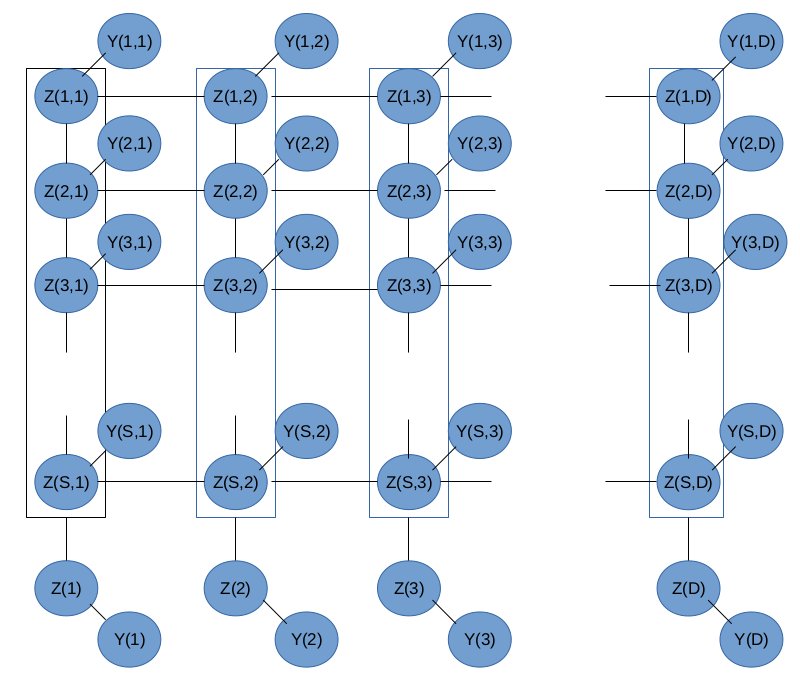}
	\caption{Proposed Markov Random Field for Anomaly Detection. Each column represents one year and each row represents one location. The horizontal edges are ``temporal edges", vertical ones are ``spatial edges", angular ones are ``data edges". For simplicity, only one or two spatial edges have been shown per location. The latent variables are shown in blue, observed ones in Green.}
\end{figure}

\subsection{Spatial and Temporal Coherence through MRF}

The spatiotemporal rainfall volume $Y_{st}$ is modeled as a multi-modal Gaussian distribution, and $Z_{st}=p$ specifies the mode (1:high,2:low,3:normal). The parameters $(\mu_{sp},\sigma_s)$ of this distribution depend on the latent state $p$ as well as location $s$, and are estimated from data. Similarly for spatial mean rainfall $Y_t$ we use a Gaussian distribution with state-specific parameters $(\mu_{p},\sigma)$. Initial estimates of these parameters can be made from the dataset using LWA to assign states.

We define \textbf{edge potential functions} so that if two vertices connected by an edge have same values of $Z$ then the corresponding edge potential is larger than if the values were different. Since the likelihood function is multiplied by these edge potentials, this encourages spatial and temporal neighbours to have same state, leading to spatial and temporal coherence. For each edge between location state node $Z_{st}$ and the corresponding AIMR state node $Z_t$ for the same year, the edge potential influences the extent to which the local state is sought to be made coherent with the aggregate state. We define potential functions for different edges as follows: 

\begin{eqnarray}
&\psi^(Z_{st},Z_{s't})= exp(C(s,s')) \textbf{ if } Z_{st}=Z_{s't}, = exp (D) \textbf{ otherwise }; \textbf{ where } s' \in NB(s)  \nonumber \\
&\psi^(Z_{st},Z_{s,t+1})= P \textbf{ if } Z_{st}=Z_{s,t+1}, = 1- P \textbf{ otherwise }; \nonumber \\
&\psi^(Z_{st},Z_{t})= exp(1/S) \textbf{ if } Z_{st}=Z_{t}, = 1 \textbf{ otherwise }; 
\end{eqnarray}

To emphasize spatial coherence, $D$ is a small constant compared to $C(s,s')$. The latter describes edge potentials if spatial neighbours are in the same state. As described previously, these edge potentials can be viewed as prior probabilities on the neighbours being in the same state. Therefore $C(s,s')$ represents a prior probability that the states in locations $s$ and $s'$ are the same, and is estimated from data. Two neighbouring grid-locations need not be highly correlated, for e.g. on either side of a narrow mountain range (such as the Western Ghats). Therefore unlike the MRF estimated by \cite{MRFdrought}, where all edges between neighbouring pairs have the same potential function, here the potentials on edges are estimated from data and are location-dependent. 

The value of edge potential $P$, for edges connecting nodes with neighbouring years, lies between zero and one. It induces temporal coherence, and hence is called the temporal coherence parameter. Higher values induce a higher emphasis on temporal coherence. 

The third set of edge potentials describes behaviour of edges between the location nodes in any given year and the AIMR-node for that year. It is defined using the exponential, so that the contribution depends on the total number of locations whose states coincide with the state assigned to the spatial mean node. $S$ is the total number of locations. The edge potential is higher when the location nodes are in the same state as the spatial mean node. 

Next, we define the \textbf{node potential functions}. These are directly proportional to the prior probabilities of the nodes being in the different states, and generally influence the threshold for anomaly detection in most real situations when data is limited and the prior is not immaterial in the MAP solution. The state that is eventually assigned in the MAP solution depends only on the relative values of these node potentials. For the default model, all node potentials are set equal to the same value, which is set to 1. But they can be varied according to the problem of interest, as described further in the Appendix.   

\textbf{MRF parameter settings:} Only the part of the likelihood function that varies with the state $Z$ affects the MAP solution. Therefore a node or edge potential can be made irrelevant to the particular analysis by making it constant, independent of the value of $Z$. In the subsequent sections, we will use this device to consider alternate settings of the MRF, including where either spatial or temporal coherence are considered in isolation.

\subsection{Anomaly Detection by Markov Random Fields}
Having defined the likelihood function, we carry out inference on the latent variables $Z$ and estimate parameters $(\mu_{sp}$, $\sigma_{s}$, $C(s,s'))$ for locations $s$, corresponding neighbours $s'$ and conditioned on latent state $p$. Unlike the maximum likelihood estimation of \cite{MRFdrought} that is based on integer programming, here we carry out inference by Gibbs Sampling, which is computationally simpler \cite{gibbsmrf}. 

Each latent variable $Z_{st}$ is initialized based on location-wise analysis described earlier, and corresponding parameters are estimated. The Gibbs sampling technique entails, at each iteration, sampling each $Z_{st}$-variable from its updated conditional distribution by conditioning on values of other variables estimated thus far in the iteration, and then re-estimating the parameters. The procedure is repeated for several iterations, and samples are collected at regular intervals. The stationary distribution of this Markov chain Monte Carlo procedure is the posterior distribution on the latent variables. The maximum a-posteriori (MAP) estimate of $Z$-variables can then be made from the samples. 

The Gibbs Sampling equation for any latent variable $Z_{st}$ or $Z_t$ is given by:
\begin{eqnarray}
p(Z_{st}=p| Z_{-s,-t},Z_t,Y_{st}) \propto p(Z_{st}=p, Z_{-s,-t},Z_t,Y_{st}) \propto p(Z_{st}=p,Z_{s't},Z_{st'},Z_t)p(Y_{st}|Z_{st}) \nonumber \\
\propto \psi^v(Z_{st}=p)\psi^f(p,Z_t)\prod_{s',t'}\psi^e(p,Z_{s't'})\mathcal{N}(\mu_{sp},\sigma_s) \nonumber \\
p(Z_{t}=q| Z_{-t},Z_{st},Y_{st}) \propto p(Z_{t}=q, Z_{-t},Z_{st},Y_{st}) \propto p(Z_{t}=q,Z_{st},Z_{t'})p(Y_{t}|Z_{t}) \nonumber \\
\propto \psi^v(Z_{t}=q)\prod_{t'}\psi^f(q,Z_{t'})\prod_{s,t}\psi^e(q,Z_{st})\mathcal{N}(\mu_{p},\sigma) \nonumber \\
\end{eqnarray}
where $s'$ refers to neighbours of $s$, $t'$ to the previous and next years, i.e. $(t-1)$ and $(t+1)$, the state $p \in \{1,2,3\}$, and $Z_{-s,-t}$ means all the $Z$-variables except $Z_{st}$. While applying this equation, we do not consider variables corresponding to spatiotemporal locations that are not neighbours of $Z_{st}$, since the Markov property of MRF holds that each node is conditionally independent of all non-neighbouring nodes conditioned on the neighbouring nodes. The Gibbs Sampling proceeds by drawing samples for each $Z_{st}$ and each $Z_t$ from Equation 3, and the optimal value for each latent variable is estimated from the distribution across these samples.

After estimating the latent-variable-set $Z$, we identify anomalies by discovering spatially and/or temporally coherent sets of spatiotemporal locations. Spatiotemporal anomalies are estimated as connected components of the MRF, such that each node of the connected component has the same value of $Z$. These values of $Z$ can be either 1 or 2, corresponding to positive and negative anomalies respectively. Due to coherence, the clusters thus identified can be at a single location but extending over several continuous years, or spatially contiguous locations in a single year, or both. Clearly, the spatio-temporal extents of the anomalies discovered this way are not fixed.

\subsection{Related Works}
Anomaly Detection is well-studied area of Data Mining~\cite{anomalysurvey}. However, its main challenge is that anomalies cannot be precisely defined, and are subjective by definition, and most papers on anomaly detection solve a specific formulation of the problem. Much of the work on anomaly detection is about classifying each individual data-point as normal or anomalous, with respect to either its immediate neighbors or the entire dataset. It is more difficult when we deal with collections of data-points rather than individually. 

While there are other approaches to anomaly detection \cite{anomalysurvey} including in case of spatiotemporal anomalies \cite{STDM}, here we use MRFs for studying coherent rainfall anomalies. MRFs themselves have been used in similar applications involving geospatial fields~\cite{MRFspatialgibbs}, including rainfall ~\cite{MRFdrought}.  Fu et al. (2012) \cite{MRFdrought} have used MRFs to detect coherent droughts of the last century, and find that their procedure can identify well-known droughts around the world. Theirs appears to be the first formulation of the rainfall anomaly detection problem in terms of MRFs. Other Bayesian models have also been considered for studying floods and droughts, such as~\cite{bayesflood} which also incorporate spatial dependence of flood properties at local scales. The present paper is partly motivated by the aforementioned work \cite{MRFdrought}. We focus on grid-level annual rainfall over India, but our method is general enough to work on rainfall data at any spatial and temporal resolution. Like \cite{MRFdrought} we use Markov Random Fields, but an important difference is that both positive and negative anomalies are considered, so that the latent variable in each node is in one of three states (positive, negative, normal). In addition the relation between anomalies at small scales (grid-wise) and large scale (all-India spatial mean) is explicitly modelled.

To identify the MAP configuration of the latent states \cite{MRFdrought} used integer programming. However, integer programming is very slow, increasing exponentially in the size of the problem, thereby necessitating probabilistic inference techniques \cite{bishop,neal}. 
In this work  we use Gibbs sampling to infer the latent variables. Gibbs sampling works by creating a Markov chain whose stationary distribution is the distribution we seek, and then carrying out a random walk on this Markov chain (\cite{mcmc,mcmc2}). Gibbs sampling has been used previously in estimating MRFs (e.g. \cite{gibbsmrf,MRFspatialgibbs}), and here we illustrate its usefulness in estimating latent states corresponding to large and heterogeneous geospatial fields such as rainfall. A survey of inference techniques for Markov Random Fields is given in~\cite{MRFgibbs}.

Geophysical spatio-temporal processes have often been studied by approaches somewhat similar to the proposed one. Models such as STARMAX~\cite{starmax} which are inspired by time-series models, express the $S$-dimensional observation vector $Y_t$ at each time-step in terms of that in the previous time-step, as $Y_t= CY_{t-1}+Dv_t+u_t$ where $u_t$ is noise, $v_t$ is input vector, and C, D are matrices that introduce spatial correlation in the elements of $Y_t$. ~\cite{seqseg} proposes an approach for temporal segmentation of multivariate time-series based on latent factors, but it is not geared for spatial coherence or anomalies. In other models such as Gaussian Random Fields~\cite{MRFgibbsfast} or Gaussian Process~\cite{GP1,GP2} the spatial correlations are more strongly captured through covariance matrices of a \emph{latent process} $X$, which is however continuous unlike our discrete process $Z$. At each time-step $t$, the observations $Y_t$ are expressed in terms of $X_t$ as $Y_t = BX_t+DV_t+u_t$, while $X$ is itself modelled with a Gaussian prior as $X_t \sim \mathcal{N}(0,\Sigma)$ or $X_t \sim GP(0,K)$ where $K$ is covariance function and $\Sigma$ is covariance matrix. As in our case, $X$ is latent and needs to be estimated conditioned on $Y$, which involves lengthy computations with the covariance matrices. Indeed a lot of research has recently investigated how such computations can be speeded up by considering covariance matrices of special forms (close to diagonal)~\cite{GP1,GP2}, or by a clever re-grouping of spatial locations which enables the $X$-variables there to be sampled simultaneously~\cite{MRFgibbsfast}. Use of discrete latent variables allow us to circumvent these issues, while providing a natural solution to anomaly detection.

\subsection{Discussion of Model}
Having discussed the model in details, and before starting its empirical evaluation, we discuss certain aspects of the model which may place it in perspective of  existing models for similar problems.

\subsubsection{Relation to other models}
Although we consider latent variables to model a spatio-temporal process, our approach is different from ~\cite{GP1,GP2,MRFgibbsfast} because we explictly want three modes - positive anomaly, negative anomaly and normal. So, a discrete latent variable serves us better than continuous. This helps us avoid the matrix computations involved in the GP-type approaches. Unlike the STARMAX-type models we do not model the temporal dynamics explicitly, nor do we ignore it as in the Gaussian Process-type models. We do not use a directed graphical model like Bayesian Networks because spatial locations cannot be ordered naturally, nor is there any known causal relation between spatial locations. So we attempt to model the joint distribution of all spatial variables instead of using conditionals as in a directed model. The spatial and temporal interactions among the $Z$ variables are modelled locally, between pairs of nodes, and the global configuration of $Z$ is inferred based on these local properties. This discrete representation used by our model is physically interpretable, and so are the local interactions. On the downside, this model is not suitable for prediction or simulation purposes for $Z$, as no conditional distribution is modelled. 

\subsubsection{Computational Complexity}
This inference process based on Gibbs Sampling is iterative, and in each iteration we need to sample the $Z$-variable for each $(s,t)$ pair, and also the $Z$-variables at all-India scale for each day. So, each iteration requires $O(ST)$ sampling steps, where $S$ is the number of locations, and $T$ the number of years. However, the sampling for each $(s,t)$ pair can be done in constant time since $Z_{st}$ can take only 3 values, and their probabilities can be computed easily based on the current $Z$-assignments to other locations. The complexity is thus linear in the number of spatio-temporal locations. Moreover, this sampling step can be sped-up by parallelized computation, where sets of $Z$-variables that are independent of each other can be sampled simultaneously. Some of these aspects have been discussed in~\cite{MRFgibbsfast}. However, a detailed study of this matter is beyond the scope of this paper.

\subsubsection{Spatio-temporal Separability}
An important issue in spatio-temporal model is that of spatio-temporal separability, i.e. whether the covariance matrix can be written as a product of a purely spatial  and a purely temporal component~\cite{spatsep}. A separable covariance matrix implies that the spatial and temporal effects can be modeled independently, which is not a good assumption in most circumstances. But in our model no such assumption is made. The covariance is a function of the edge potentials, and the covariance between the $Y$-variables at a pair of spatio-temporal locations $Y(s,t)$ and $Y(s',t')$ can be written as a product of all edge potentials along the graph path between these two nodes, through  $Z(s,t)$ and $Z(s',t')$ along the spatial and temporal edges joining them (see Fig 1), marginalizing over the $Z$-variables on this path. The sum-product form of this term, along with the form of the edge-potential functions, ensures that spatial and temporal effects are not separable in this model, which is a good assumption for spatio-temporal data. Since the latent space $Z$ being modelled here is discrete rather than continuous, we are able to avoid making separability assumptions without using complex computations, as discussed in~\cite{spatsep}. 

\section{Test of Method}
We fit the MRF model discussed above, and also perform the location-wise analysis (LWA) discussed previously in Section 2.1 on a dataset of $1^{\circ}-1^{\circ}$ gridded rainfall data measured all over India, for the period 1901-2011. This grid system has 357 locations over India ($S=357$). The data is available at daily scale, but for the analysis in this paper we compute annual aggregate values. The $Z$-values are computed and anomalies are discovered. Before going into details of spatio-temporal properties, we first provide a test of the method, by reproducing some known results about AIMR. The results from the MRF are compared with LWA to highlight the differences and benefits.

\subsection{Local Anomalies in given years} 
We examine results from two years: 1998 (declared excess-rainfall year by IMD) and 2002 (declared deficient-rainfall year). Maps of positive and negative anomalies in these two years are shown in Figure 2. The first panel in each pair shows results from the LWA, while the second panel shows those of the MRF. Overall the maps have many similarities, which seem to validate the MRF approach. 

It was noted previously that LWA may yield isolated anomalies as well, and this is seen in the figure. By contrast, the constraint of spatial coherence in the MRF yields more spatially connected and extensive anomalies, with fewer isolated anomalies. Anomalies of both kinds are more spatially contiguous with the MRF. 

Furthermore, for the excess rainfall year, the MRF yields a larger number of locations with positive anomaly state compared to LWA (84 in MRF as compared to 75 in LWA). Likewise, in the deficit-rainfall year, the MRF yields more locations having negative anomalies (194 in MRF compared to 147 in LWA). This is a result of the edges connecting the location-specific states $Z_{st}$ to the aggregate state $Z_t$ in the MRF, which have higher edge potential when the corresponding nodes are in the same state, as well as the effects of spatial coherence. 

\begin{figure}
	\centering
	\includegraphics[width=10cm,height=3cm]{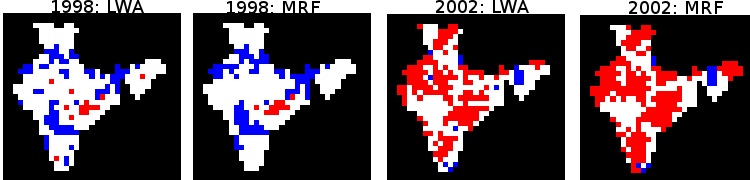}
	\caption{Comparison of results of MRF with location-wise analysis (LWA) for 1998 (excess-rainfall year) and 2002 (deficient-rainfall year). First panel in each pair shows results for LWA. Colors indicate different latent-states (blue: positive; red: negative; white: normal). In case of MRF, anomalies are more spatially contiguous.}
\end{figure}

\subsection{AIMR anomalies and local anomalies} 
The variables $Z_t$ denote the anomaly corresponding to All-India spatial mean rainfall (AIMR). For each year we compute AIMR $Y_t$ from local measurements $\{Y_{st}\}$, and from this time-series estimate mean $\mu$ and standard deviation $\sigma$ across years. The excess rainfall years $H$ are defined as those with $Y_{t} \geq \mu+\sigma$ and deficient-rainfall years $L$ have $Y_{t} \leq \mu-\sigma$.  

These definitions do not depend on how widespread are local anomalies but only on amount of spatial mean rainfall. We can instead define all-India anomalies so as to depend on the widespread occurrence of local anomalies. For any year $t$, we compute the number of locations under anomalies of either kind ($N1(t)$ and $N2(t)$) as found by LWA, and corresponding means ($\mu_{N1}$, $\mu_{N2}$) and standard deviations ($\sigma_{N1}$, $\sigma_{N2}$) across the years. Based on these, we identify those years with exceptionally large numbers of locations under  positive anomalies $(HL)$ and exceptionally large numbers of locations under negative anomalies $(LL)$. In other words, $HL = |t: N1(t) \geq \mu_{N1}+\sigma_{N1}|$ and $LL = |t: N2(t) \geq \mu_{N2}+\sigma_{N2}|$. It turns out that $H$ and $HL$ are not equal, and their overlap $\frac{|H\cap HL|}{|H|}$ is only 0.7. Similarly $L$ and $LL$ are also not equal, and $\frac{|L\cap LL|}{|L|}$ is only 0.7. This illustrates that the aggregate state $Z_t$ when defined based on spatial mean rainfall often takes different values from when it is defined based on widespread occurrence of local anomalies.

In the MRF model, edge potentials ensure that assignment of $Z_t$ is also influenced by values of the location-wise latent states $Z_{st}$, and large numbers of local anomalies of one kind increase the probability of $Z_t$ being assigned to the same anomaly. At the same time, it also takes into account the AIMR estimate $Y_t$. Hence in the MRF the value of $Z_t$ should be able to capture all kinds of all-India anomalies defined so far - $H,HL,L,LL$. 

Let $ZH$ and $ZL$ be the positive and negative years identified by the MRF, i.e. $ZH=\{t: Z_t=1\}$ and $ZL=\{t: Z_t=2\}$. The set $ZH$ captures very well the contents of both $H$ and $HL$, with $\frac{|H\cap ZH|}{|H|}=1$ and $\frac{|HL\cap ZH|}{|HL|}=0.92$. Similarly $ZL$ also overlaps well with $L$ and $LL$, with $\frac{|L\cap ZL|}{|L|}=1$ and $\frac{|LL\cap ZL|}{|LL|}=0.84$. This shows that the MRF model helps discover both types of all-India anomalies, based on spatial-mean rainfall as well as widespread occurrence of local anomalies, simultaneously.

We describe anomaly statistics from the MRF for extreme years ($H,L$), where all-India rainfall is either excess or deficient.  Generally, across approaches it can be expected that in years of $H$ (excess rainfall) the number of locations (N1H) assigned as positive state $Z_{st}=1$ is much higher than positive state locations in all other years (N1Y), while the number of locations (N2L) assigned to negative state in $L$ (deficit rainfall years) is much higher than in all other years (N2Y). These relationships are seen for the MRF with spatial coherence and LWA in Table 2. 

Spatial coherence in the MRF causes the mean number of nodes with positive state in years of excess rainfall to be higher than in case of LWA (Table 2). Similarly there are more negative state assignments in years of deficit rainfall as compared to LWA. Furthermore, the mean difference between number of locations with positive and negative states in H and L years respectively (D12H, D21L) is more pronounced with the MRF than in case of location wise analysis (Table 2). Spatial coherence favours occurrence of the corresponding anomaly states in excess or deficit rainfall years. 

Thus the proposed approach links AIMR states to local states, which helps to identify extreme-rainfall years in a more inclusive way. It also helps to localize the anomalies formed in such years.

\begin{table}
	\centering
	\begin{tabular}{| c | c | c | c | c | c | c |}
		\hline
		Method & N1Y & N2Y & N1H & N2L & D12H & D21L \\
		\hline
		LWA     & 54 & 54 & 107 & 111 & 101  & 86  \\
		MRF	  & 62 & 58 & 132 & 129 & 118  & 103 \\
		\hline
	\end{tabular}\caption{Mean number of spatial locations under positive (1) and negative (2) states in all years (N1Y,N2Y), only excess-rain (H) years and only deficient-rain (L) years (N1H,N2L). Also, difference (D12H,D21L) between number of nodes with positive and negative states in H and L years. Results are shown for the MRF and location-wise analysis (LWA). Compared to LWA, spatial coherence in the MRF increases occurrence of corresponding local positive and negative states in excess and deficit rainfall years.}
\end{table}

\section{Effects of MRF edge potentials}
Clearly, the assignment of the latent state variables $Z$ at the different spatio-temporal locations is strongly influenced by the edge potentials of the MRF. It can be generally expected that many isolated locations that are assigned to states 1 or 2 by LWA, will be assigned to state 3 by MRF to preserve spatial coherence. On the other hand, some locations assigned to state 3 by LWA may be assigned to states 1 or 2 if several of their neighbors are in such states. In this section, we study how the $Z$-assignments are affected by different parameter settings of the MRF involving the edge potentials.

\subsection{Assignment Statistics}

For different parameter settings, we compute the total number of nodes in the entire graph assigned states 1 and 2 ($N1$, $N2$). We also compute \emph{confusion matrices} to describe the degree of overlap between anomaly nodes found by location-wise analysis and the MRF. $NG1$ denotes the number of positive state nodes ``gained" by the proposed method when compared to LWA, i.e. nodes satisfying $Z_{st}=1, Z0_{st}\neq 1$ (Recall that state assignments by LWA are $Z0$). These are nodes not part of positive anomalies by LWA, but part of positive anomalies in the corresponding MRF. Similarly $NL1$ is the number of positive state nodes ``lost" by the proposed method compared to LWA, i.e. nodes satisfying $Z0_{st}=1, Z_{st}\neq 1$. The number of negative state nodes ``gained" and ``lost" in this way are denoted as $NG2$ and $NL2$ respectively.

\subsection{Edge Potential Settings}

First we isolate effects of spatial coherence, in the absence of temporal coherence. Absence of temporal coherence is implemented by using constant edge potentials for all edges across years. We also use constant node potentials for all nodes and states.

Next we study the effects of temporal coherence alone, leaving out spatial coherence effects.  We consider effects of temporal coherence with parameter $P$ ($MRF-TC-P$), with increasing $P$ denoting increasing emphasis on temporal coherence.  The node potential is uniform, independent of the assignment of latent variable $Z$.  Results are shown in Table 2 and Figure 3. 

In the presence of temporal coherence, the number of nodes in positive state is much larger than that of nodes in negative state. The relative difference increases as the temporal coherence parameter increases. 

As the role of temporal coherence is increased, by increasing $P$ from 0.5 to 0.99, the number of anomaly-states decreases. Increasing coherence generally leads to fewer anomaly-states. That is why it is not possible to generalize the effect of MRF compared to LWA, without also specifying the coherence parameters.

In general the number of anomaly-states ``lost" when switching from LWA to the MRF is higher as either spatial or temporal coherence is introduced, and as the temporal coherence parameter is increased. This is expected, as many anomalies found by LWA are isolated and do not reflect coherent effects on larger scales. A less expected effect of introducing coherence is that a significant number of new anomalies are ``gained", i.e. identified when LWA could not extract them. Such anomalies are manifested at larger scales only.

Finally we consider the MRF where both spatial coherence and temporal coherence, the latter having parameter $P$, are present ($MRF-STC-P$). In the presence of spatial coherence, the effects of increasing the temporal coherence parameter $P$ are similar to the previous discussion in the context of temporal coherence alone: higher temporal coherence parameter leads to fewer anomaly-states. Furthermore, the number of positive states is larger than the number of negative states, and the relative difference becomes larger as temporal coherence is increased. 

There can be different approaches to enforcing spatial coherence based on Equation 2, and we consider the effects in the following. We contrast five different approaches, for which the results are shown in the last part of Table 2. For this analysis, $P$ is kept at 0.9.

In the first three cases, $D=0$. That is, the edge potentials for spatial neighbours have zero weight if the latent states differ. These approaches differ in the choice of edge potentials $C(s,s')$ between spatial neighbours in case the latent states are the same:  ``prop", where for neighbouring pairs of locations, $C(s,s')$ is proportional to the number of years that the locations have the same phase i.e. sign of rainfall change; ``anml"  where for neighbouring pairs of locations, $C(s,s')$ is proportional to the number of years that the locations had the same state as estimated by LWA; and ``unif" where for neighbouring pairs of locations, $C(s,s')$ values are equal. An important result is that these three approaches do not have much effect on statistics of state assignments(Table 2). Therefore anomaly detection using MRFs does not depend much on details of the spatial coherence model as long as the edge potentials in the presence of spatial coherence are much higher than edge potentials when the neighbouring states differ; recall that for these three cases $D=0$ so that ratio $C/D$ is infinity. 

In the last two approaches towards spatial coherence, we relax the constraint that $D=0$. This is essentially a weakening of the spatial coherence requirement. The ratio of $C$ and $D$ can, however, affect the relative weight given to spatial coherence, with higher ratios emphasizing spatial coherence more. We consider two settings: ``mxd1" where $C=2, D=1$ and ``mxd2" where $C=5, D=1$. If ratio $C/D$ is higher, there are fewer anomaly nodes (Table 2). The ``prob" setting with $D=0$ has been used for all the analysis done before and after this analysis.

\begin{table}
	\centering
	\begin{tabular}{| c | c | c | c | c | c | c |}
		\hline
		Method      & N1   & N2 & NG1 & NG2& NL1 & NL2 \\
		\hline
		LWA         & 5666 & 5621 & -    &  -  &  -   &  -  \\
		MRF-SC      & 6561 & 6038 &  481 & 678 & 1376 & 1095\\
		\hline
		MRF-TC-0.50 & 7905 & 7645 & 2248 & 2031&    9 &    7\\
		MRF-TC-0.75 & 6687 & 6379 & 1319 & 1079&  298 &  321\\
		MRF-TC-0.90 & 5482 & 4725 & 1065 & 725 & 1249 & 1621\\
		MRF-TC-0.99 & 3555 & 2178 &  910 & 408 & 3028 & 3844\\
		\hline
		MRF-STC-0.50& 6484 & 6049 & 1313 &1090 &  495 &  663\\
		MRF-STC-0.75& 4916 & 4322 &  583 & 410 & 1333 & 1709\\
		MRF-STC-0.90& 3447 & 2282 &  361 & 185 & 2580 & 3524\\
		MRF-STC-0.99& 1828 & 1105 &  204 &  96 & 4042 & 4612\\
		\hline 
		MRF-STC-unif& 1755 & 1013 &  192 &  83 & 4103 & 4691\\
		MRF-STC-prop& 1828 & 1105 &  204 &  96 & 4042 & 4612\\
		MRF-STC-anml& 1808 & 1109 &  196 & 113 & 4054 & 4625\\
		MRF-STC-mxd1& 3125 & 1785 &  704 & 276 & 3252 & 4105\\
		MRF-STC-mxd2& 2200 & 1328 &  379 & 166 & 3934 & 4597\\
		\hline
	\end{tabular}\caption{Total number of nodes assigned to the different states of the entire graph, in different settings of MRF, and the number of nodes under anomaly-states ``gained" and ``lost" compared to location-wise analysis (LWA). See Sections 4.1, 4.2 for notations. Increasing the temporal coherence parameter decreases the number of nodes in anomaly-states. Increasing the temporal coherence parameter makes nodes under positive state more predominant. Increasing the ratio of $C$ and $D$ in the spatial coherence model leads to fewer nodes under anomaly-states.}
\end{table}

\begin{figure}
	\centering
	\includegraphics[width=8cm,height=4cm]{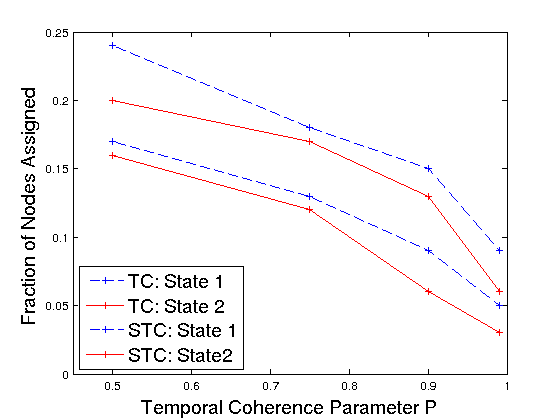}
	\caption{Fraction of spatiotemporal locations assigned to both anomaly states in different settings of MRF: using only temporal coherence and using both spatial and temporal coherence}
\end{figure}

\section{Properties of Discovered Anomalies} 
In Section 2.5, we discussed how the local state variables $Z$ assigned by MRF or LWA are used to identify spatio-temporally coherent zones as positive or negative anomalies. In this section we study the properties of these anomalies, under the different settings of the MRF discussed in the previous section.

An important question is how widespread and persistent positive and negative rainfall anomalies are. Another important question is, how much different the rainfall volumes are from the long-term climatology, in case of each anomaly. To evaluate these, we first define several properties of the anomalies. 

\subsection{Anomaly Statistics}

The \emph{spatiotemporal size} of each anomaly is the size of the corresponding connected component in the graph, i.e. the number of nodes present in it. We measure the STS: mean spatiotemporal size of all anomalies, including all years; and similarly the STSP: mean spatiotemporal size of all positive anomalies; and STSN: mean spatiotemporal size of all negative anomalies. We define the \emph{spatial size} of an anomaly as the number of distinct spatial locations included in the nodes covered by it. The \emph{temporal size} of an anomaly is similarly defined as the number of distinct years included in it. We thereby estimate mean spatial size of all anomalies (SS), only positive (SSP) and only negative (SSN) anomalies. Similarly we measure (TS, TSP, TSN) for corresponding mean temporal sizes. 

Each state of $Z$ at each location is associated with a distribution over rainfall values. Fig 4 shows the mean rainfall values for each of the locations and each state of $Z$. Mathematically, these are $mean_{t: Z_{st}=1}(Y_{st})$ for positive anomalies, and $mean_{t: Z_{st}=2}(Y_{st})$ for negative anomalies. Two different settings of the MRF are considered: using spatio-temporal coherence with temporal coherence parameters $P=0.7$ and $P=0.9$, and the ``prop" setting of spatial coherence. The plots show that these mean rainfall fractions for the different states (shown by green, blue and red plots) are clearly well-separated in most locations.

To quantify the \emph{severity} or ``anomalousness" of each anomaly quantitatively, we first compute the ratio of the rainfall received at each spatio-temporal location covered by the anomaly, and the long-term mean rainfall over each of these locations. We define the \emph{intensity parameter} of the anomaly as the mean of these ratios. Mathematically, let $A$ be the set of spatio-temporal locations affected by a particular positive or negative anomaly $a$. For each $(s,t) \in A$, we compute $F_a(s,t)=\frac{Y_{st}}{\mu_s}$. Then the intensity of anomaly $a$ is given by $I_a=mean_{(s,t)\in A}F_a(s,t)$.

\begin{figure}
	\centering
	\includegraphics[width=5cm,height=3cm]{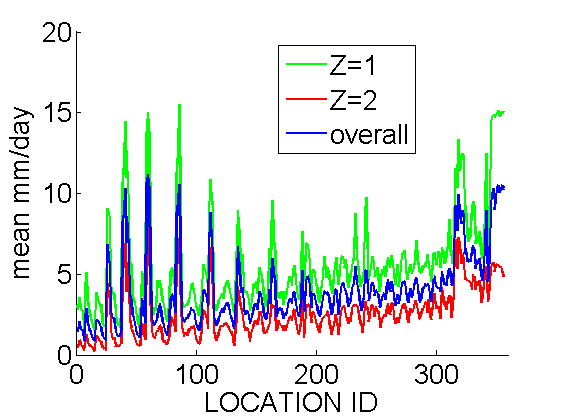}\includegraphics[width=5cm,height=3cm]{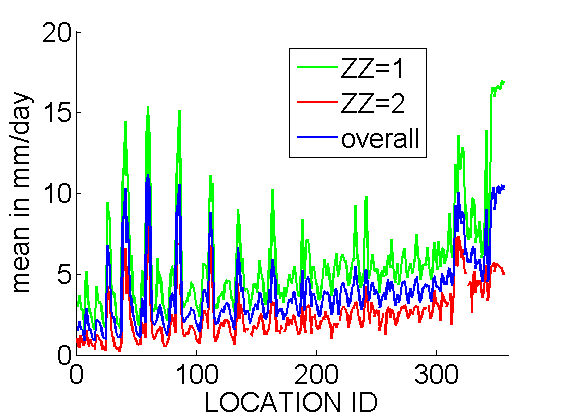} 			   
	\caption{The mean rainfall at each of the locations in the two anomaly states, and overall, for MRF settings using $P=0.7$ (left) and $P=0.9$ (right), along with spatial coherence.}
\end{figure}

\subsection{Effect of MRF settings}

We consider location-wise analysis (LWA), and using MRFs under different settings. These settings include only spatial coherence (SC), only temporal coherence with parameter $P$ ($TC-P$) and both spatial and temporal coherence ($STC-P$). Results are shown in Table 3. The different groups of columns show the number of anomalies, spatiotemporal size, spatial size, temporal size and intensity respectively, each one separately for positive and negative anomalies. 

The results indicate complex relationships involving spatial and temporal scales of anomalies. As expected, with LWA, the number of anomalies is much larger and their mean sizes much smaller, in comparison to versions of the MRF where various constraints of coherence are present. 
In the absence of spatial coherence, as the temporal coherence parameter is increased, the spatial size of anomalies becomes smaller. Larger temporal coherence parameter selects for more long-lived anomalies and hence these tend to become smaller in spatial extent. The spatiotemporal size decreases as the temporal coherence parameter is increased. The aforementioned effect is also present when spatial coherence is included in the MRF. The selection for longer but spatially less extended anomalies when the temporal coherence parameter is increased creates a trade-off between spatial and temporal extents. Such a trade-off is intrinsic to spatio-temporal anomaly detection: with a larger emphasis on a certain type of coherence (spatial or temporal) the corresponding size of anomalies increases while the other size decreases. 

In Table 3, we also study the mean intensity of the anomalies under different settings of MRF. Clearly, as either type of coherence is increased, the mean intensity parameter of positive anomalies increases, and that of negative anomalies decreases, and given the aforementioned definition of this parameter the selected anomalies are more ``intense". This is a welcome result, indicating that use of spatio-temporal coherence helps us to identify severe anomalies, rejecting mild ones.

\begin{table}
	\centering
	\begin{tabular}{| c | c | c || c | c || c | c || c | c || c | c |}
		\hline
		& \multicolumn{2}{|c||}{\#Anomalies} & \multicolumn{2}{|c||}{S-T size} & \multicolumn{2}{|c||}{Spatial sizes} & \multicolumn{2}{|c||}{Temporal sizes} & \multicolumn{2}{|c||}{Intensity}\\            
		\hline
		Method      & NP   & NN   & STSP & STSN &  SSP   & SSN  & TSP  & TSN & IP  & IN \\
		\hline
		LWA         & 1085  & 1163 &  5.3 & 5.0  &  5.0   &  4.4 & 1.1  & 1.2 & 1.34 & 0.70\\
		\hline         
		MRF-SC      & 519   & 472  & 11.5 & 12.5 & 10.8   & 11.0 & 1.1  & 1.2 & 1.37 & 0.69\\
		\hline
		MRF-TC-0.50	& 1083  & 1155 &  6.9 &  6.6 &  6.3   & 5.8  & 1.2  & 1.2 & 1.24 & 0.80\\
		MRF-TC-0.75	& 1000  & 1105 &  6.7 &  6.1 &  5.7   & 5.1  & 1.3  & 1.2 & 1.26 & 0.78\\
		MRF-TC-0.90	& 795   &  825 &  6.5 &  5.9 &  4.8   & 5.0  & 1.6  & 1.5 & 1.30 & 0.76\\
		MRF-TC-0.99	& 472   &  365 &  6.7 &  6.8 &  3.4   & 2.8  & 2.3  & 2.6 & 1.34 & 0.73\\  
		\hline
		MRF-STC-0.50& 550   &  459 & 10.8 & 12.6 & 10.0   & 11.1 & 1.2  & 1.2 & 1.32 & 0.75\\
		MRF-STC-0.75& 401   &  317 & 11.5 & 13.5 & 10.0   & 11.0 & 1.3  & 1.3 & 1.37 & 0.71\\
		MRF-STC-0.90& 303   &  137 &  9.5 & 15.3 &  7.5   &  9.3 & 1.4  & 1.8 & 1.44 & 0.68\\
		MRF-STC-0.99& 208   &   75 &  7.0 & 15.8 &  3.9   &  6.4 & 1.9  & 2.7 & 1.47 & 0.67\\
		\hline
	\end{tabular}\caption{Mean spatial, temporal, spatiotemporal sizes and mean intensities of positive and negative anomalies in different settings of edge potentials of MRF. A trade-off between the spatial and temporal sizes of anomalies is inherent to anomaly detection; and illustrated here by varying the temporal coherence parameter. Spatial coherence effect in the MRF leads to larger spatial size of detected anomalies, which correspondingly have shorter mean temporal size. Larger temporal coherence parameter leads to longer mean temporal size and correspondingly smaller mean spatial size. Also, the anomalies become more intense (high intensity for positive and low intensity for negative) as the spatio-temporal coherence are increased.}
\end{table}

\begin{table}
	\centering
	\caption{Correlations between different pairs of statistics for individual positive and negative anomalies, computed by different methods}
	\begin{tabular}{| c || c | c || c | c || c | c || c | c |}
		\hline
		& \multicolumn{2}{|c||}{Temp.size} & \multicolumn{2}{|c||}{Spat-temp.size} & \multicolumn{2}{|c||}{Spat-temp.size} & \multicolumn{2}{|c||}{Spat-temp size}\\
		& \multicolumn{2}{|c||}{Spat.size} & \multicolumn{2}{|c||}{Spat.size} & \multicolumn{2}{|c||}{Temp.size} & \multicolumn{2}{|c||}{Intensity}\\
		\hline
		Method       & Pos   & Neg   & Pos   & Neg   & Pos   & Neg   & Pos   & Neg\\   
		\hline
		LWA          & 0.42  & 0.40  & 0.99  & 0.94  & 0.49  & 0.61  & 0.15  & -0.1\\
		\hline 
		MRF-SC       & 0.43  & 0.33  & 0.99  & 0.93  & 0.51  & 0.61  & 0.01  & 0.1\\
		\hline
		MRF-TC-0.5   & 0.45  & 0.44  & 0.99  & 0.96  & 0.50  & 0.60  & 0.23  & -0.2\\
		MRF-TC-0.7   & 0.43  & 0.44  & 0.98  & 0.96  & 0.52  & 0.59  & 0.19  & -0.2\\
		MRF-TC-0.9   & 0.38  & 0.37  & 0.95  & 0.82  & 0.57  & 0.69  & 0.18  & -0.1\\
		MRF-TC-0.99  & 0.33  & 0.29  & 0.80  & 0.70  & 0.69  & 0.65  & 0.06  & -0.1\\
		\hline
		MRF-STC-0.5  & 0.40  & 0.26  & 0.99  & 0.94  & 0.48  & 0.53  & 0.14  & -0.2\\
		MRF-STC-0.7  & 0.38  & 0.31  & 0.97  & 0.87  & 0.52  & 0.68  & 0.06  & -0.1\\
		MRF-STC-0.9  & 0.34  & 0.18  & 0.94  & 0.79  & 0.57  & 0.67  & 0     & 0\\
		MRF-STC-0.99 & 0.32  & 0.22  & 0.87  & 0.65  & 0.60  & 0.76  & -0.1  & 0\\
		\hline
	\end{tabular}
\end{table}

\begin{figure}
	\centering
	\includegraphics[width=6cm,height=4cm]{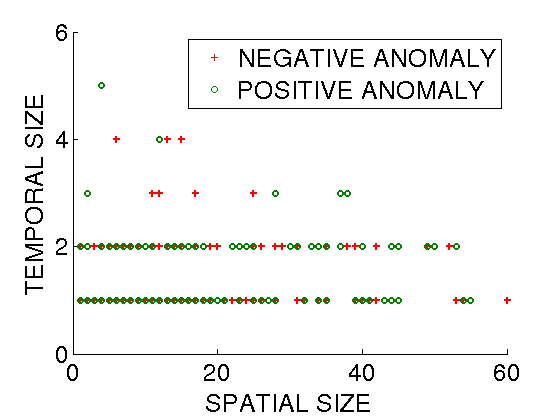}\includegraphics[width=6cm,height=4cm]{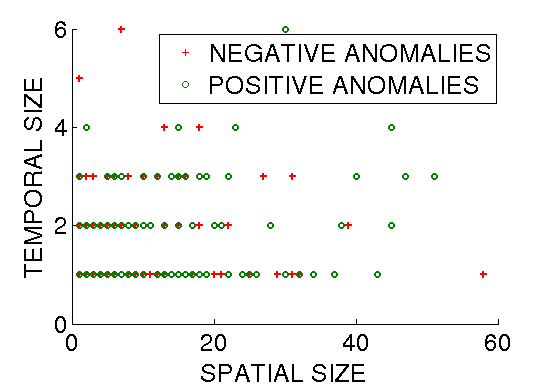}
	\includegraphics[width=6cm,height=4cm]{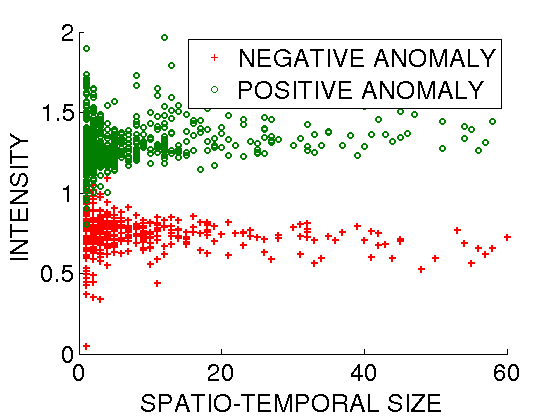}\includegraphics[width=6cm,height=4cm]{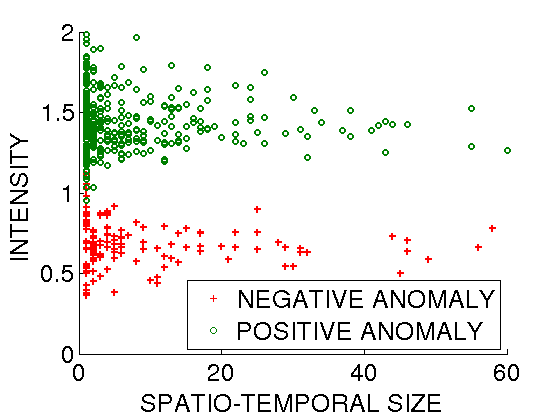}
	\caption{Above: Temporal versus spatial sizes of individual positive and negative anomalies, in fixed parameter settings. Spatial coherence (prop) is used with two choices of the temporal coherence parameter (left: $P=0.50$, right: $P=0.90$). Larger anomalies tend to be shorter-lived, but there are individual exceptions and large variability exists in the sizes of individual anomalies. Below: Spatio-temporal size versus intensities of the same set of anomalies.}
\end{figure}

\subsection{Variations among Anomalies}

The above discussion pertained to parameter-based tradeoffs in mean spatial and temporal sizes of anomalies. However, even for fixed parameter settings of the MRF, there is substantial variation in size and intensity of the detected anomalies. Such variation of spatial and temporal sizes is shown in Figure 5 for two realizations of the MRF. It is seen that generally larger anomalies tend to be shorter-lived, but there are individual exceptions. There is a large range of temporal sizes for a known spatial size, for both positive and negative anomalies. In Figure 5 we also plot the variation of intensity with spatio-temporal size of the anomalies in two realizations of MRF. Here the correlation is even weaker.

We compute the correlations between these statistics of individual anomalies. Once again, this is done separately for each setting of the MRF considered in Table 3, and separately for positive and negative anomalies. The results are shown in Table 4. It shows that in almost all the settings the correlation between spatio-temporal size and spatial size is very strong, though it reduces as the temporal coherence parameter $P$ is increased (i.e. the mean temporal size of the anomalies increase). The correlation between spatio-temporal and temporal sizes is less strong, though it increases slightly with $P$. The spatial and temporal sizes are less well correlated. There is no noticeable correlation between spatio-temporal size and intensity.

\section{Case studies of some Anomalies}

Next, we investigate some of the anomalies individually, which were discovered using MRF with spatio-temporal coherence, with temporal coherence parameter $P=0.9$. 

We first consider a positive anomaly that occurred in the states of Odisha and Jharkhand along the eastern coast (Fig 6A), in the year 1994. This anomaly covered 20 grid-locations, but persisted for only 1 year (spatial size 20, temporal size 1). The long-term mean annual rainfall over the concerned 20 grid-locations is 4.18 mm per day per location, but that year the mean rainfall over these locations was 5.84 mm per day per location (anomaly intensity of $1.4$). Overall, the year 1994 was classified as a positive anomaly year in terms of AIMR, with mean rainfall of 4.23 mm per day per location, compared to the long-term mean of 3.94 mm per day per location (intensity of $1.1$). The map of locations having local positive and negative anomalies in 1994 are shown in Fig 6B, which indicates that the Odisha anomaly was quite significant. The LWA-based local anomalies are shown in Fig 6C. Another major anomaly occurred roughly in the same area (Fig 6D) in 2001, covering 11 locations. The mean rainfall that year over this anomaly was 5.5 mm per location per day, compared to the long-term mean of 4.1 mm per location per day (intensity of $1.3$). The year 2001 was classified as normal at all-India scale, and the map in Fig 6E shows the locations under positive and negative anomalies according to MRF. 

A significant negative anomaly occurred around a stretch of Central India (Fig 7A) in 2000, which was classified as an all-India negative anomaly year. The anomaly map by MRF of the year is shown in Fig 7B. The anomaly covered 22 locations which receive 3.88 mm per day per location rainfall on average, but in that year they received only 2.17 mm (anomaly intensity of $0.56$). Again, around 10 locations in Odisha near the eastern coast (Fig 7D) had a negative anomaly in 2002, which was a major drought year in terms of AIMR. These locations, which receive 4.18 mm on average, received only 2.4 mm in 2002 (intensity of $0.57$). The MRF-based anomaly map for 2002 is shown in Fig 7E, while Fig 7F shows the local anomalies by LWA. 

Some anomalies are temporally extended, i.e. they cover several years. A good example is a positive anomaly that covered 5 years from 1987 to 1991, over the Meghalaya and Southern Assam region, covering 24 locations (Fig 8A). The mean annual rainfall over these locations is 6.35 mm per location per day, but in these 5 years, the mean rainfall volumes were 6.99, 8.53, 6.92, 7.14 and 7.45 mm per location, per day. Among these years, only 1988 and 1990 were classified as positive anomaly at all-India scale, while the other three years were classified as normal. The MRF-based anomaly map of 1987 is shown in Fig 8B. Again, 11 locations in the south-western state of Kerala (Fig 8D), one of the wettest parts of India, suffered a negative anomaly stretching over 1985-87, all of which were classified as normal years. The mean rainfall over these locations is 6.15 mm per location per day, but during these three years, this mean was 4.83, 4.59 and 4.9 respectively. The MRF-based anomaly map of 1985 is shown in Fig 8E.

\begin{figure}
	\centering
	\includegraphics[width=3.5cm,height=3cm]{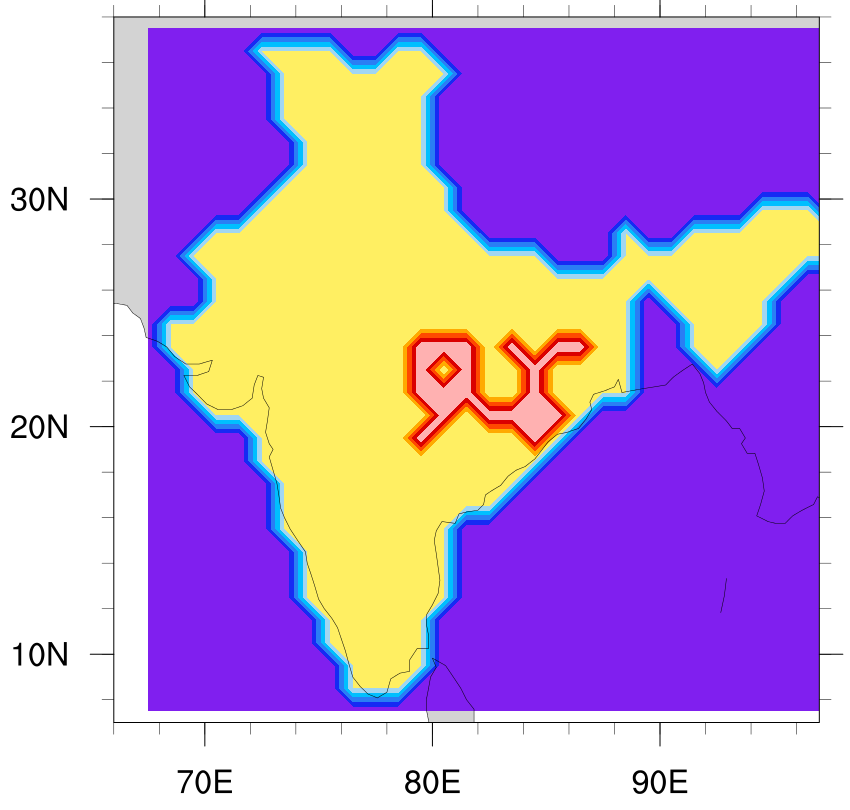}\includegraphics[width=3.5cm,height=3cm]{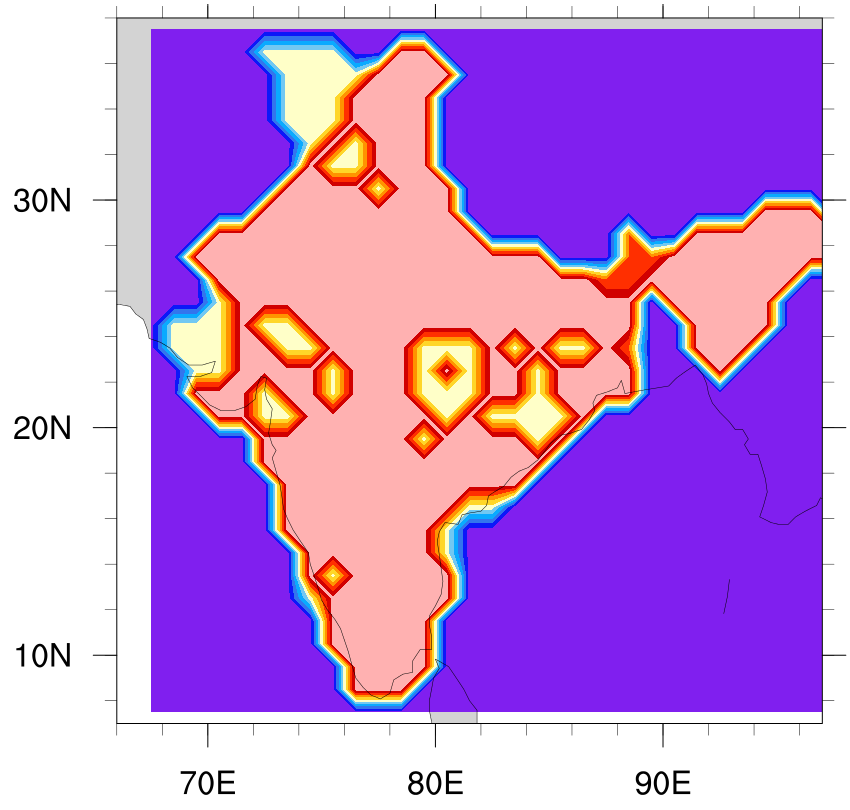}\includegraphics[width=3.5cm,height=3cm]{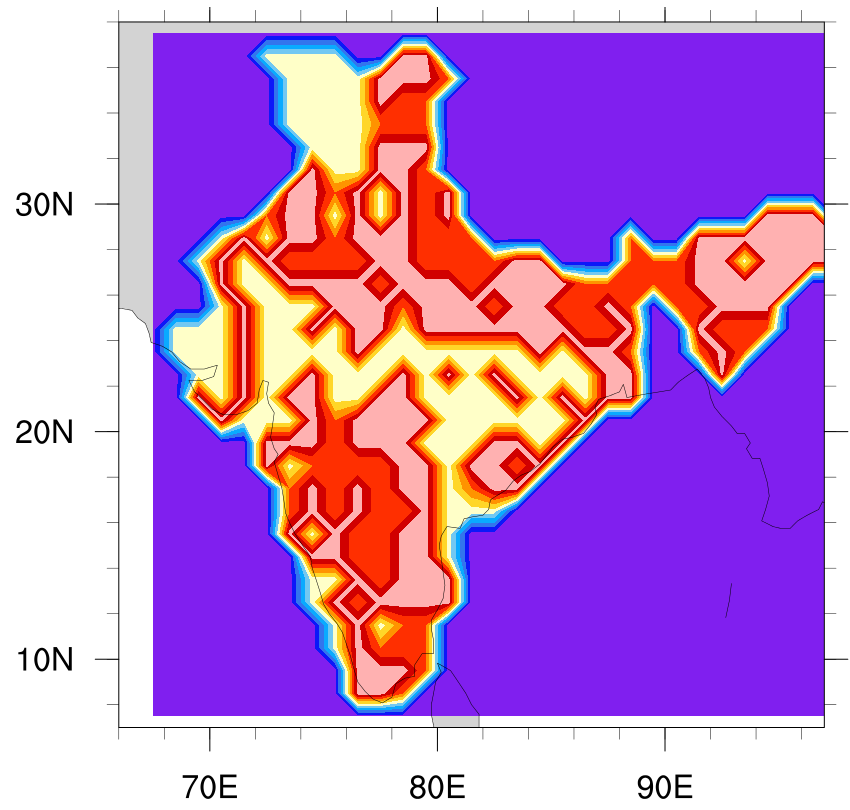} 		   
	\includegraphics[width=3.5cm,height=3cm]{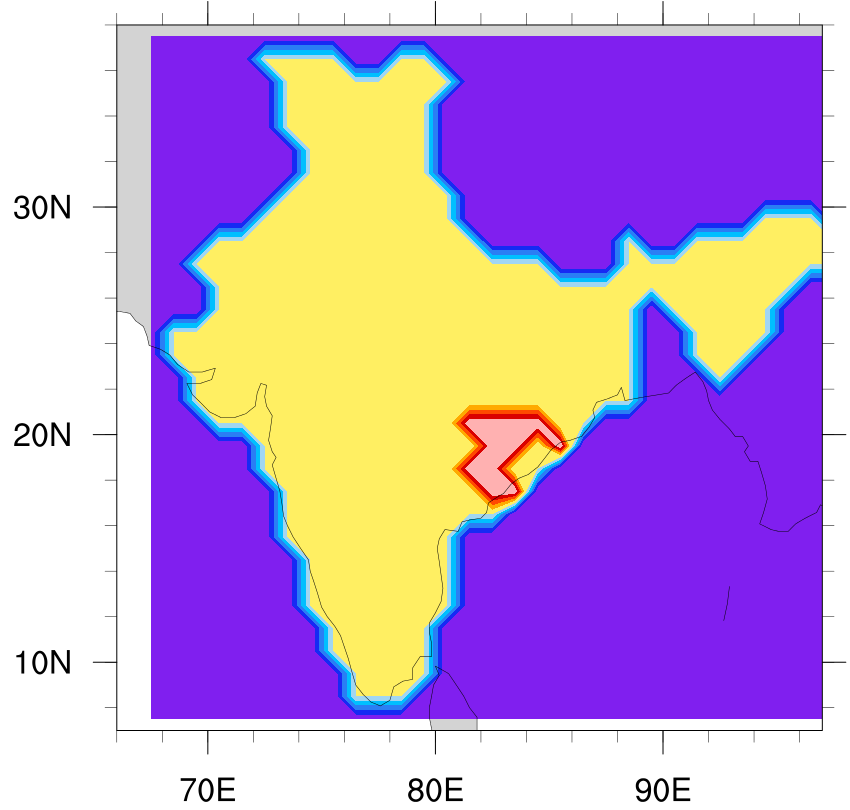}\includegraphics[width=3.5cm,height=3cm]{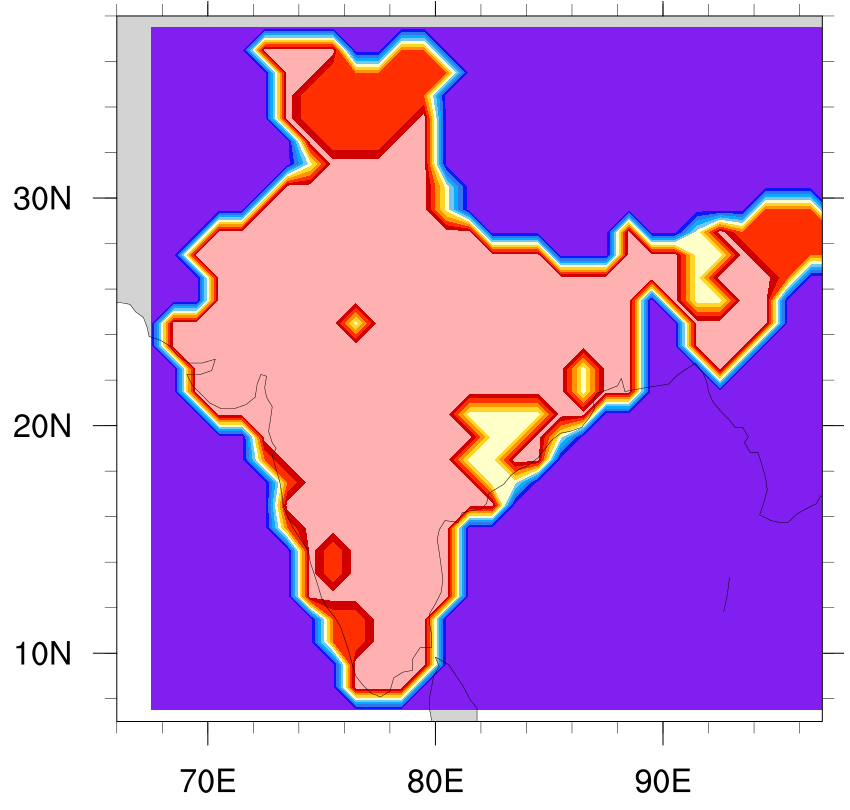}\includegraphics[width=3.5cm,height=3cm]{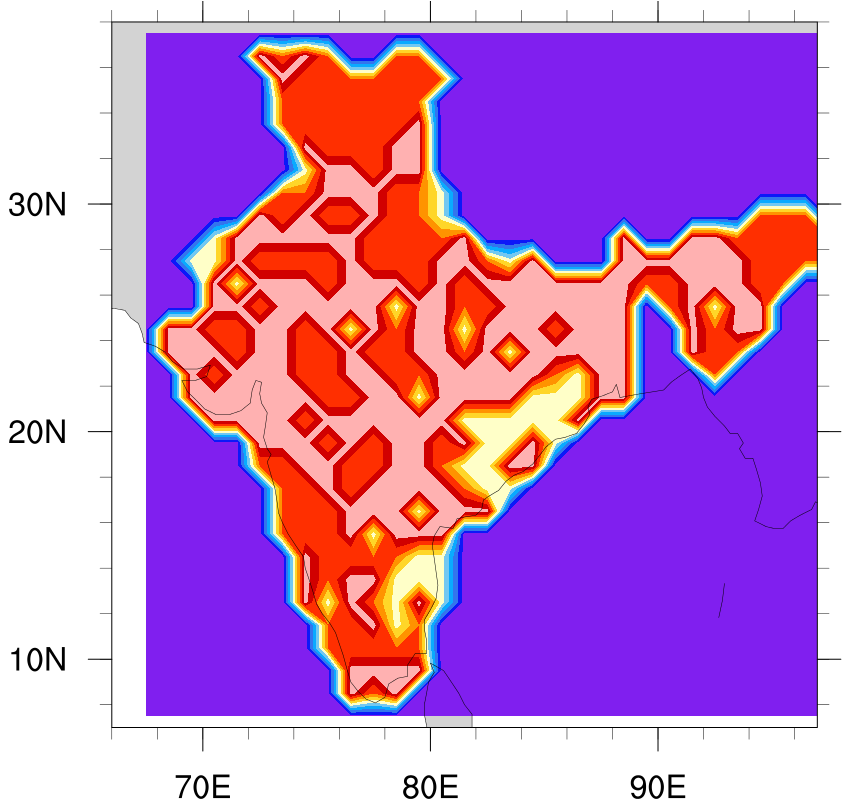}
	\caption{ABOVE: 6A: a set of locations that formed a positive anomaly in 1994, shown in pink. 6B: MRF-based Anomaly map for 1994 (yellow: positive, red: negative). 6C: LWA-based local rainfall anomaly map for 1994.  BELOW: 6D: a set of locations that formed a positive anomaly in 2001, shown in pink. 6E: MRF-based Anomaly map for 2001 (yellow: positive, red: negative). 6F: LWA-based local rainfall anomaly map for 2001.}
\end{figure}

\begin{figure}
	\centering
	\includegraphics[width=3.5cm,height=3cm]{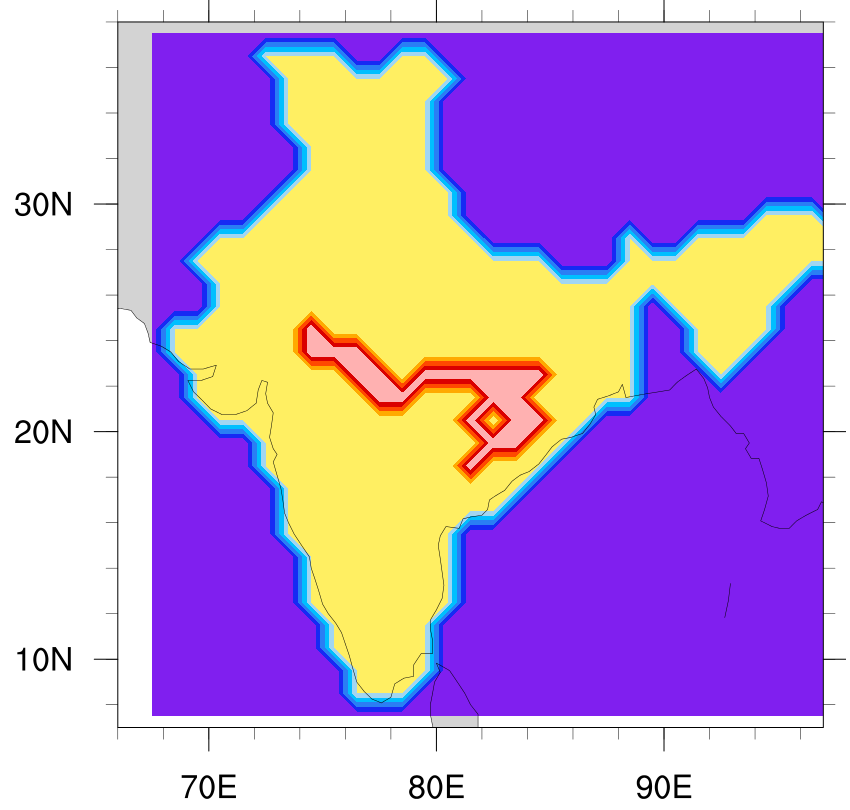}\includegraphics[width=3.5cm,height=3cm]{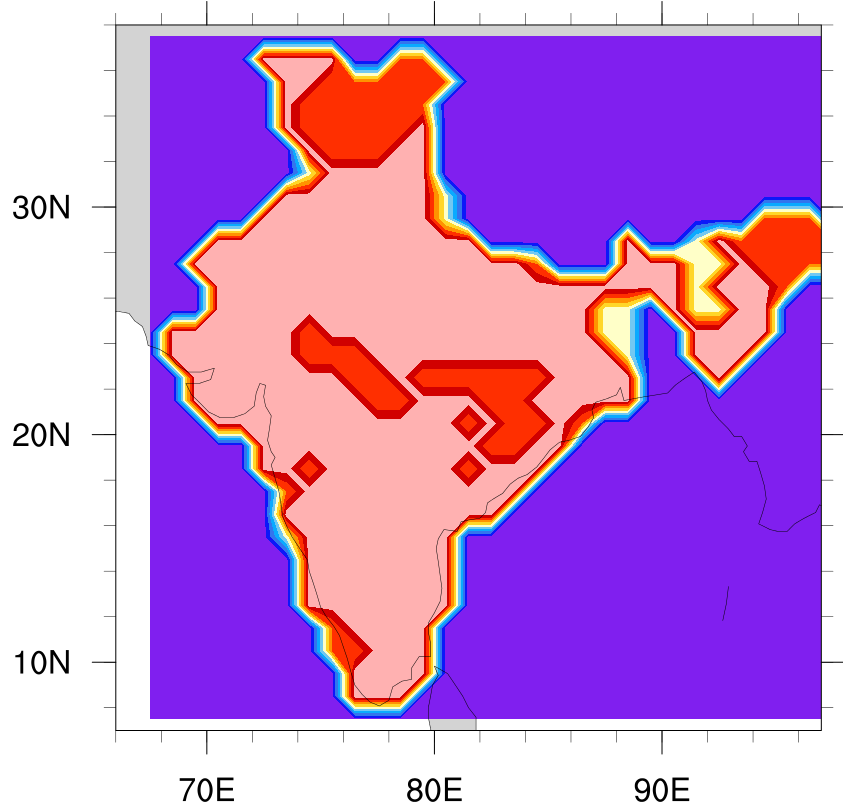}\includegraphics[width=3.5cm,height=3cm]{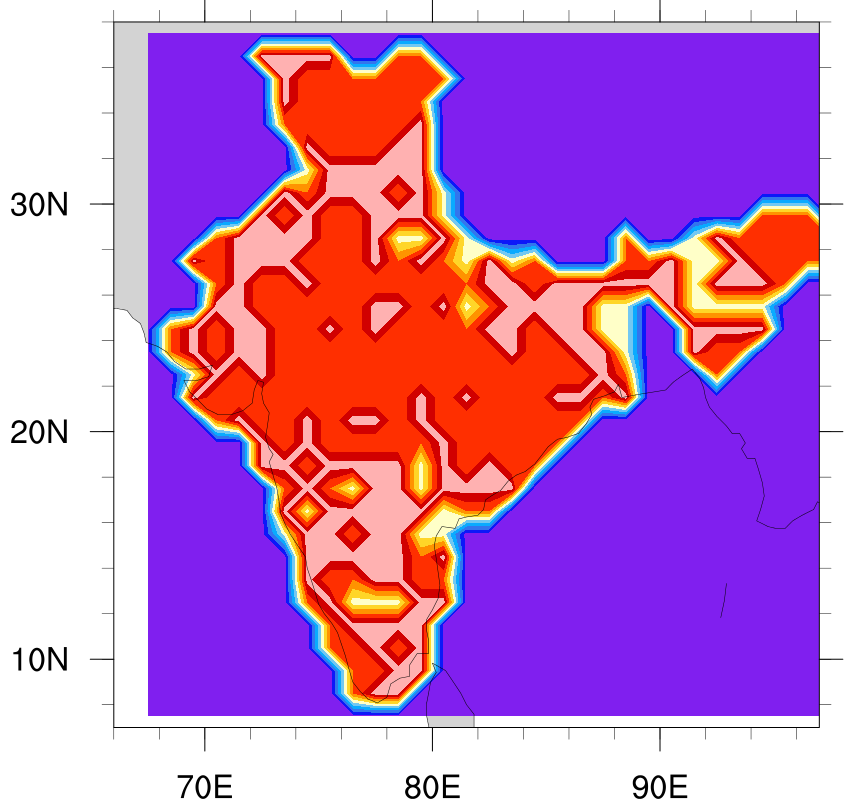}		   
	\includegraphics[width=3.5cm,height=3cm]{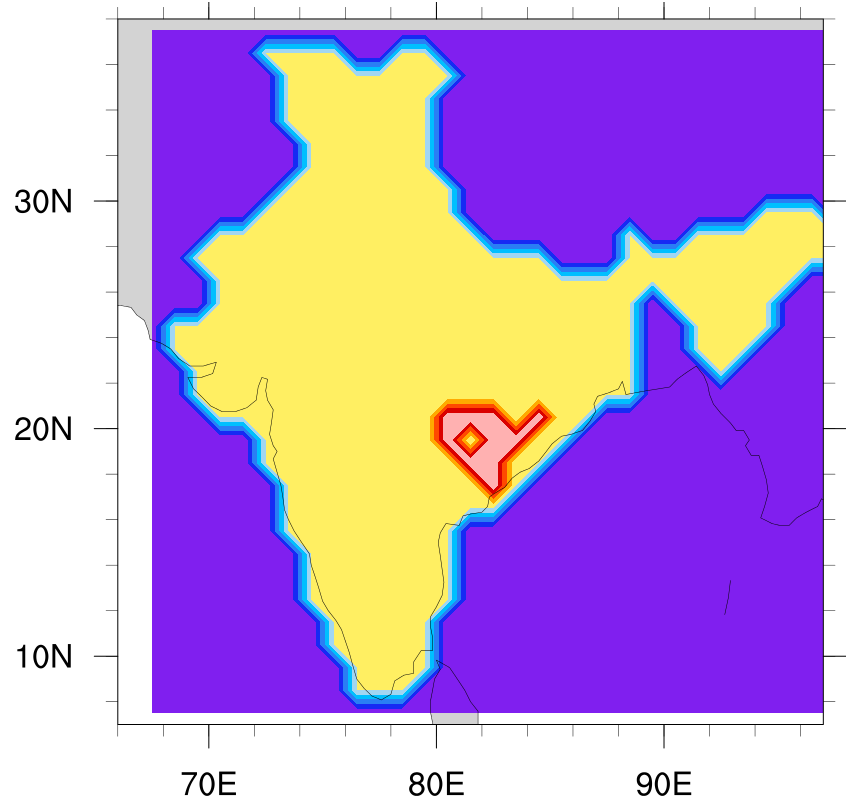}\includegraphics[width=3.5cm,height=3cm]{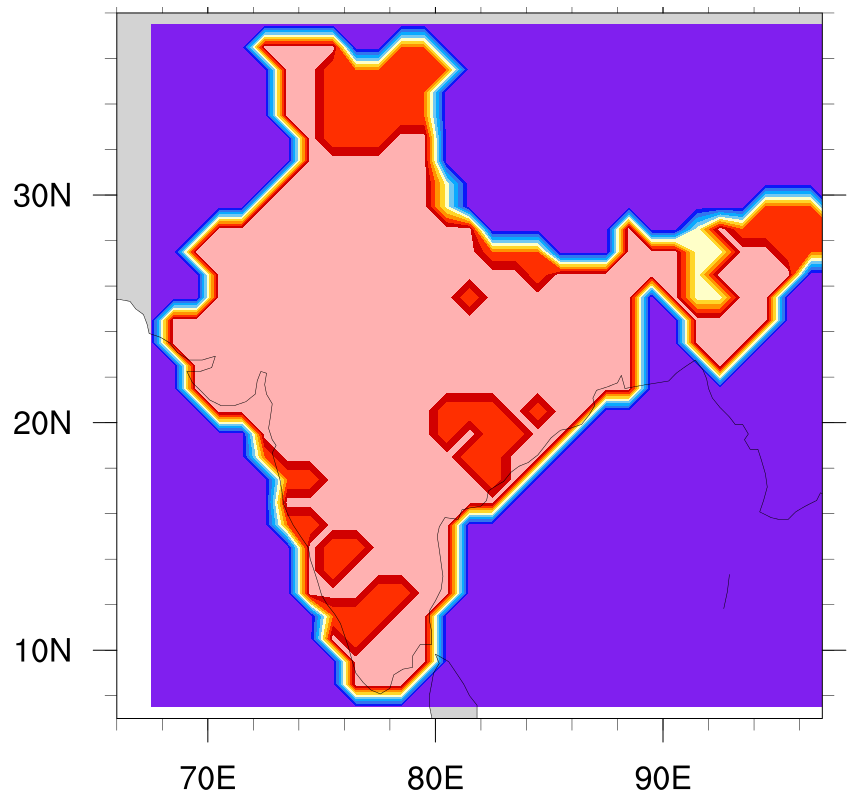}\includegraphics[width=3.5cm,height=3cm]{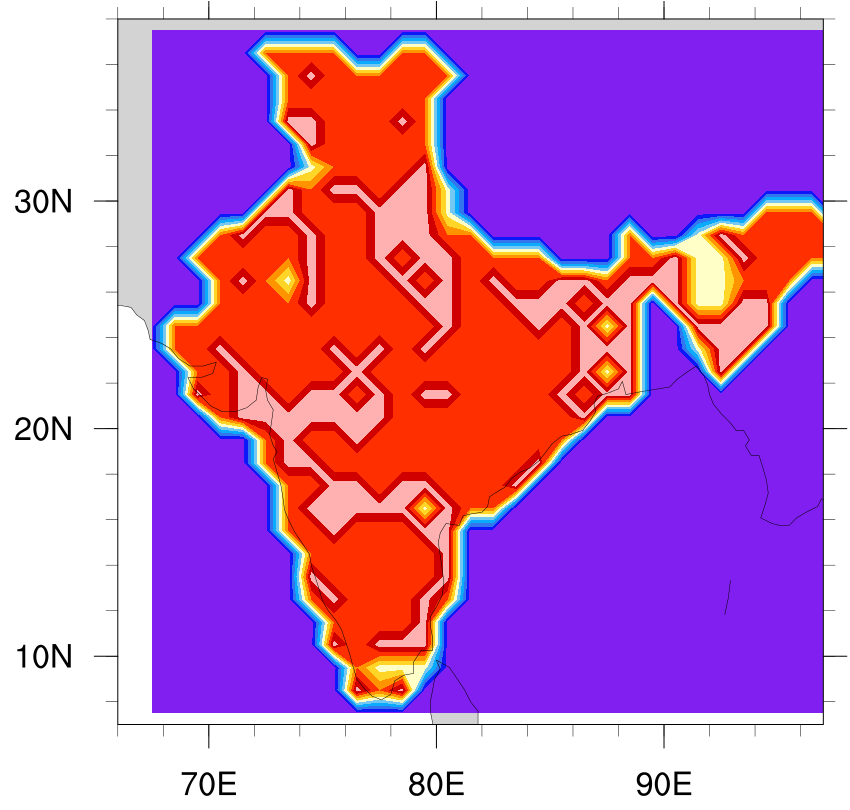}
	\caption{ABOVE: 7A: a set of locations that formed a negative anomaly in 2000, shown in pink. 7B: MRF-based Anomaly map for 2000 (yellow: positive, red: negative). 7C: LWA-based local rainfall anomaly map for 2000.  BELOW: 7D: a set of locations that formed a negative anomaly in 2002, shown in pink. 7E: MRF-based Anomaly map for 2002 (yellow: positive, red: negative). 7F: LWA-based local rainfall anomaly map for 2002.}
\end{figure}

\begin{figure}
	\centering
	\includegraphics[width=3.5cm,height=3cm]{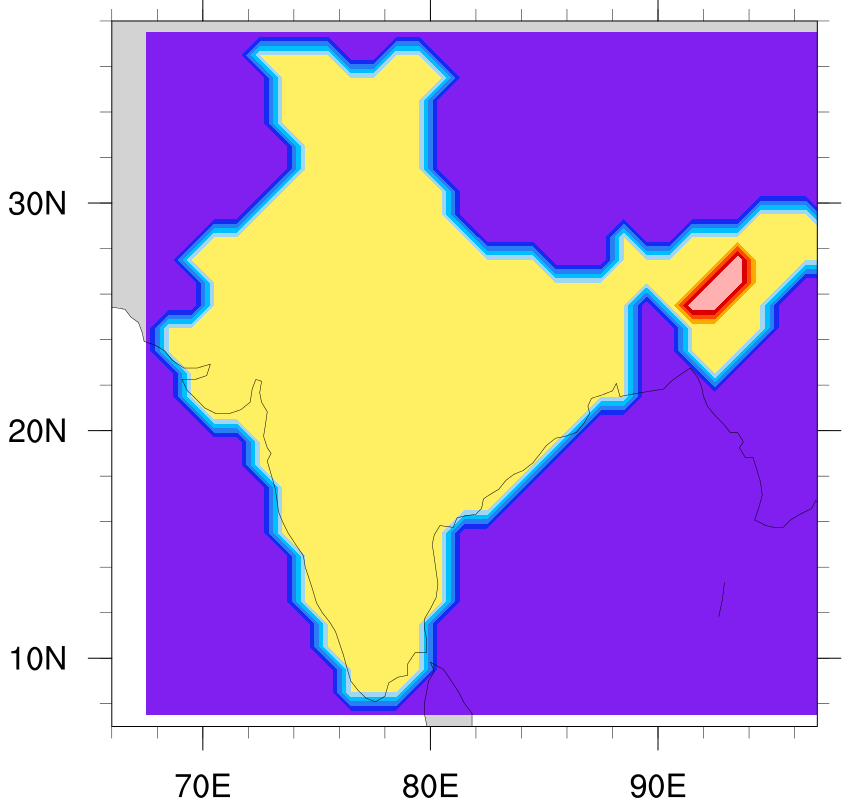}\includegraphics[width=3.5cm,height=3cm]{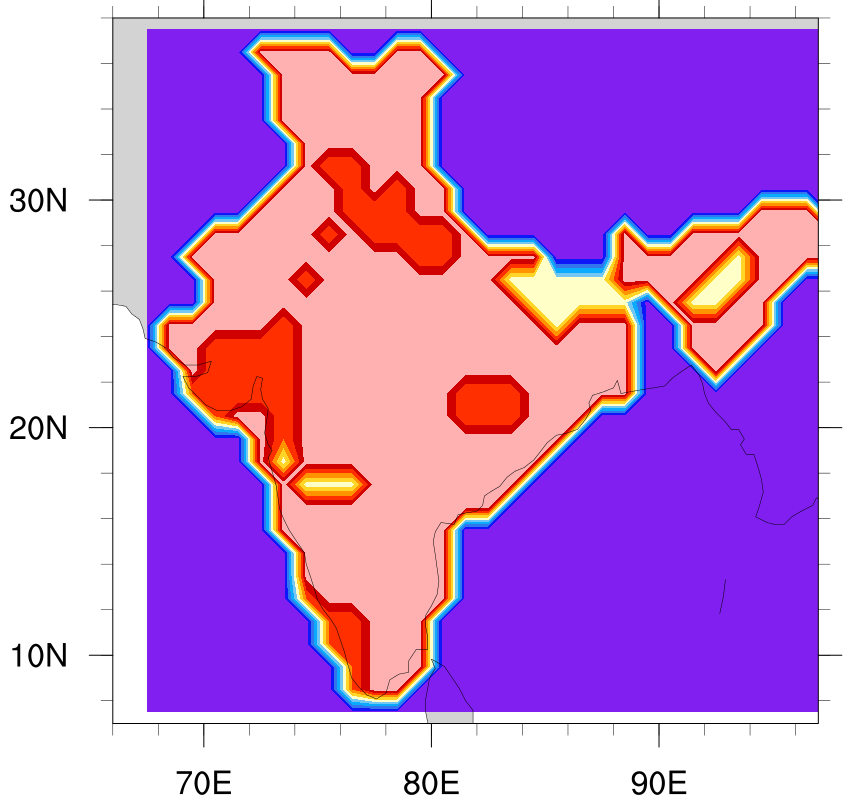}\includegraphics[width=3.5cm,height=3cm]{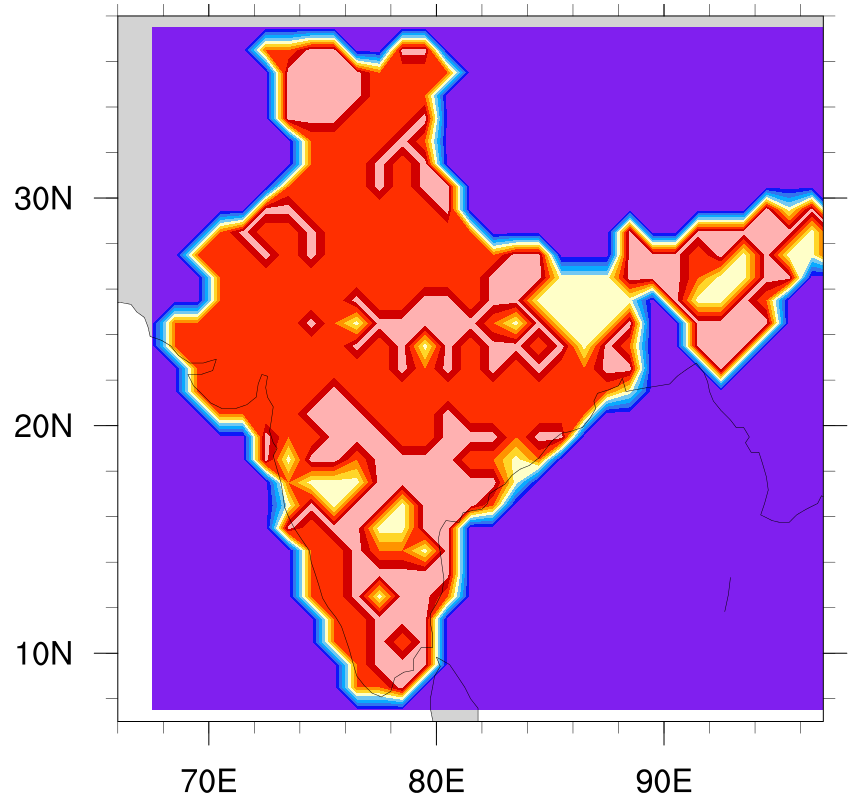}
	\includegraphics[width=3.5cm,height=3cm]{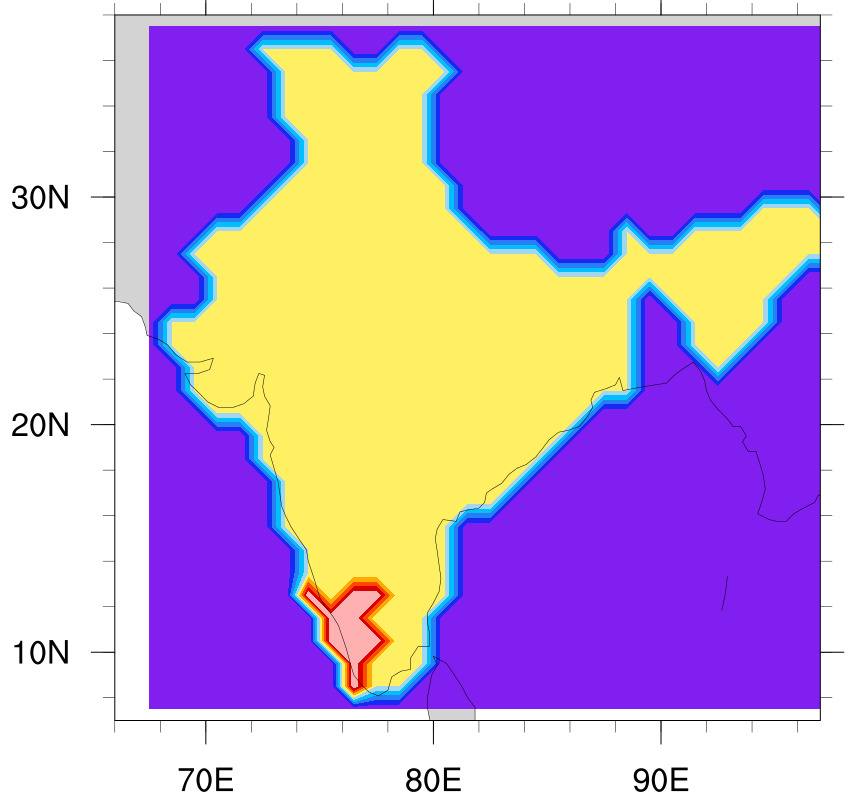}\includegraphics[width=3.5cm,height=3cm]{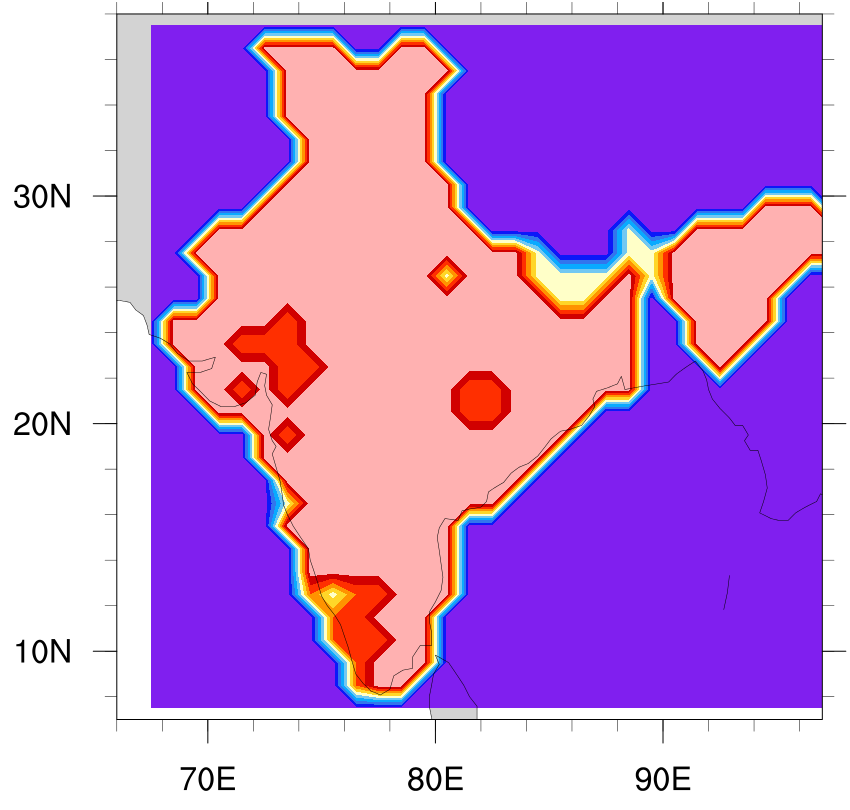}\includegraphics[width=3.5cm,height=3cm]{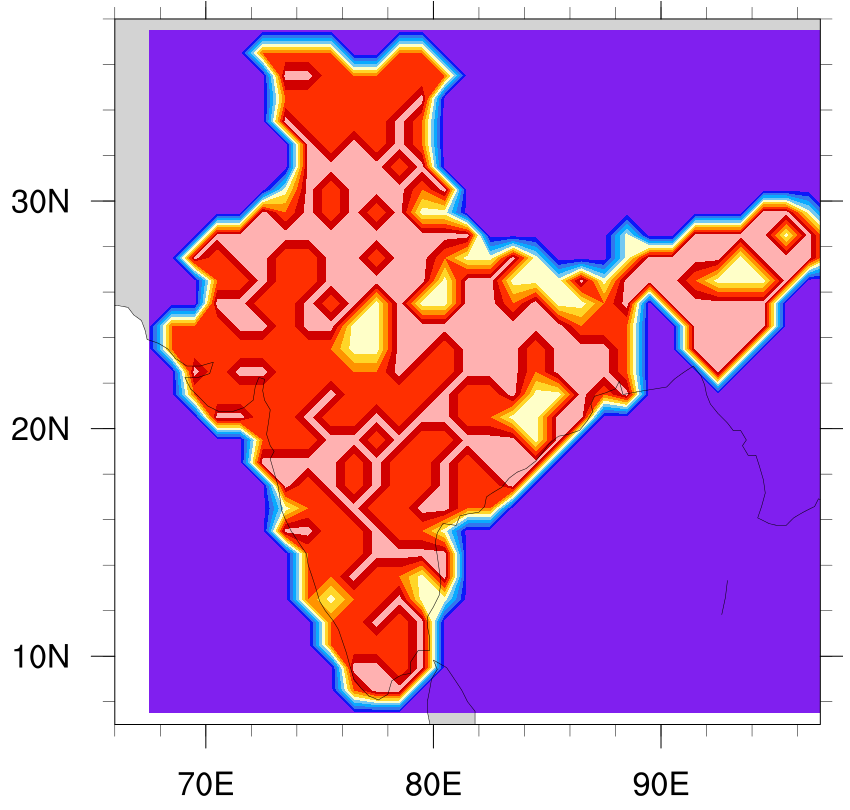}
	\caption{ABOVE: 8A: a set of locations that formed a positive anomaly in 1987-91, shown in pink. 8B: MRF-based Anomaly map in 1987 (yellow: positive, red: negative). 8C: LWA-based local rainfall anomaly map for 1987. BELOW: 8D: a set of locations that formed a negative anomaly in 1985-87, shown in pink. 8E: MRF-based Anomaly map in 1985 (yellow: positive, red: negative). 8F: MRF-based Anomaly map in 1985}
\end{figure}

\section{Conclusions}
This paper describes a method for coherent anomaly detection using Markov Random Fields (MRFs), where each node is associated with a location and year. Coherence is emphasized because it is an inherent property of rainfall, and also because anomalies are consequential especially when extended spatially or temporally. The anomaly states are represented as latent random variables, so probabilistic methods are required for their estimation. For this purpose we use Gibbs sampling, a type of Markov chain Monte Carlo method. We also consider sensitivities of the results to parameters of the MRF. 

The MRF is able to identify more coherent anomalies compared to traditional analysis using location-specific thresholds. MRFs offer a principled approach to handling the heterogeneity and anisotropy in the occurrence of anomalies, where more traditional methods such as wavelets may not be appropriate. The method can discover intense positive and negative anomalies of various sizes, without requiring any thresholds. Furthermore the method can be used to characterize both the occurrence of anomalies at large spatial scale by assigning a state variable for All-India spatial-mean rainfall, as well as the widespread occurrence of grid-scale anomalies through effects of edge potentials and spatial coherence in the MRF. 

The effects of edge potentials enforcing coherence as well as node potentials influencing the threshold for anomaly detection within the MRF are described. We show that adjusting the parameters has effects that are consistent with intuition. However the results are not overly sensitive to the parameters. One effect of coherence is to reveal anomaly states that are classified as normal in location-wise threshold-based analysis, because of the influence of neighbouring locations being assigned to anomaly states. Increasing spatial coherence through edge potentials leads to fewer but larger anomalies. Enforcing any one type of coherence more strongly, selects for either longer-lived or spatially more extended anomalies, though fewer in number. On the other hand, increasing spatio-temporal coherence results in selection of more ``intense" anomalies  instead of mild ones.

There is also variability in the spatial and temporal sizes of anomalies. Anomalies longer in one dimension (spatial/temporal) tend to be shorter in the other. Furthermore positive anomalies are not necessarily larger or smaller than negative anomalies, as the results vary with choice of parameters. 

Overall, this study provides some understanding of heterogeneities in rainfall over Indian region. The results also raise the question of whether the anomalies discovered by this method are relevant for understanding hydrological floods and droughts, which are based on considering multiple variables, including soil moisture. A natural extension of this work would be to infer anomaly states based on the inclusion of additional climatic and hydrological variables. 

Clearly, anomalies are a very significant feature of rainfall in general and Indian rainfall in particular, and any realistic simulation of regional rainfall should be able to capture their salient properties. Statistics of coherent anomalies learnt from MRF-based approaches could present further tests and benchmarks of regional-scale rainfall simulations made from climate models and statistical simulators. 

\section{Acknowledgments}This research was supported by Airbus India Postdoctoral Fellowship and Divecha Centre for Climate Change, Indian Institute of Science. We are thankful to Dr. J. Srinivasan and Dr. V.Venugopal for valuable inputs. 
\bibliography{references}

\section{Appendix}

\subsection{Choice of Node Potentials}

As described in Section 2.4, node potentials can be varied depending on the problem being considered.  These potentials can be viewed as prior probabilities on the occurrence of different states. 

For example, a lower threshold on anomaly detection is achieved by specifying $\psi^v({Z_{st}=1}) = C_1$,  $\psi^v({Z_{st}=2}) = C_2$ and $\psi^v({Z_{st}=3}) = C_3$, where $C1$ and $C2$ are high while $C3$ is low. Relative frequencies of positive and negative anomalies can be adjusted by changing $C_1$ and $C_2$ accordingly. Another application might be to vary node potentials by location. In locations receiving low average rainfall ($\mu_s$ is small), negative anomalies may be more consequential and hence important to detect. Likewise, locations receiving higher average rainfall ($\mu_s$ is high) might be more sensitive to flooding events. We define the set of locations receiving low average rainfall as $L$ and those receiving high average rainfall as $H$. Then
\begin{eqnarray}	
&\psi^v({Z_{st}=1}) = C_1,  \psi^v({Z_{st}=2}) = C_2 \textbf{ and } \psi^v({Z_{st}=3}) = C_2& \textbf{ when } s \in L	\nonumber \\
&\psi^v({Z_{st}=1}) = C_2,  \psi^v({Z_{st}=2}) = C_1 \textbf{ and } \psi^v({Z_{st}=3}) = C_2& \textbf{ when } s \in H \nonumber \\
&\psi^v({Z_{st}=1}) = C_3,  \psi^v({Z_{st}=2}) = C_3 \textbf{ and } \psi^v({Z_{st}=3}) = C_3& \textbf{ in other locations} \nonumber \\
\end{eqnarray}

To achieve the above, we specify $C_1 \leq C_2$. On the contrary, the goal may be to identify positive anomalies in dry locations, or negative anomalies in wet locations, by specifying $C_2 \leq C_1$. 

Yet another application may involve inducing homogeneity of heterogeneity in anomaly detection, by identifying positive anomalies especially during years of strong mean rainfall or negative anomalies in the reverse situation respectively. Alternatively, the objective may be to identify negative anomaly states during dry years or vice versa. For this type of problem, we denote sets of years with excess and deficient spatial mean rainfall as $H$ and $L$. Once again defining node potentials as
\begin{eqnarray}	
&\psi^v({Z_{st}=1}) = C_1,  \psi^v({Z_{st}=2}) = C_2 \textbf{ and } \psi^v({Z_{st}=3}) = C_2& \textbf{ when } t \in L \nonumber \\
&\psi^v({Z_{st}=1}) = C_2,  \psi^v({Z_{st}=2}) = C_1 \textbf{ and } \psi^v({Z_{st}=3}) = C_2& \textbf{ when } t \in H\nonumber \\
&\psi^v({Z_{st}=1}) = C_3,  \psi^v({Z_{st}=2}) = C_3 \textbf{ and } \psi^v({Z_{st}=3}) = C_3& \textbf{ in other years } \nonumber \\
\end{eqnarray}

Homogeneity can be achieved by specifying $C_1$ to be low and $C_2$ high, and heterogeneity with the reverse specifications. There is clear analogy between the two sets of problems, one in which node potentials are adjusted by location and the second where the type of year is the primary factor.

\subsection{Effects of node potentials} 
Node potentials influence the thresholds for anomaly detection, and can be interpreted as prior probabilities of the corresponding anomaly being present before any observations are made. To examine the effects we compute the \emph{mean number} of positive (NP) and negative (NN) anomalies with spatiotemporal size above 1. In all cases, we maintain spatial and temporal coherence through edge potentials in the MRF, with temporal coherence parameter $P=0.99$. 

In setting NP1, we consider equal weights for all 3 states at each node; NP2 favours detection of positive anomalies by setting $C_1=2, C_2=1, C_3=1$; NP3 favours negative anomalies by setting $C_1=1, C_2=2, C_3=1$ in NP3; NP4 prioritizes both anomalies over the normal state using $C_1=2, C_2=2, C_3=1$. 

One might also set node-specific potentials depending on statistics at either the location or the year associated with the node. We define set LS of dry locations, where mean annual rainfall $(\mu_s)$ is atleast one standard deviation $\sigma$ below the mean of this quantity across locations $(\mu)$, i.e. $LS=\{s: \mu_s \leq \mu-\sigma\}$. We also define set HS of wet locations, where $HS=\{s: \mu_s \geq \mu+\sigma\}$. 

In NP5 we set node potentials $C_1=2, C_2=1$ in nodes of HS, and $C_1=1, C_2=2$ in nodes of LS. This favours positive anomalies in wet locations, and negative anomalies in dry locations. In contrast, the values are reversed in NP6, favouring positive anomalies in dry locations and negative anomalies in wet locations. 

For introducing year-specific node potentials, we consider deficient-rain years L and excess-rain years H once again. In NP7 we set $C_1=2, C_2=1$ in nodes of H, and $C_1=1, C_2=2$ in nodes of L. This favours positive anomalies in excess-rain years and negative anomalies in deficit-rain years. These settings are reversed in NP8, favouring positive anomalies in deficit-rain years and negative anomalies in excess-rain years. 

Table 5 shows anomaly statistics for the various settings of node potentials examined here. When giving additional weight to positive anomalies (as in cases NP2, NP4) the number of positive anomalies increases as would be expected. Similarly when negative anomalies are given higher weight (as in cases NP2, NP4) the number of negative anomalies increases. A common tendency across these settings is that the number of distinct positive anomalies is much larger than that of negative anomalies, but negative anomalies have larger mean spatiotemporal size. 

Emphasizing node-specific potentials that depend on features of either the location or the year associated with the node, in NP5-NP8, does not substantially change the overall statistics, but affects the particular anomalies detected (which are not shown). In NP7, where in AIMR anomaly years the local anomalies of the same type are favoured, the difference between mean sizes of negative and positive anomalies decreases. This is mainly because positive anomalies have higher spatial size than negative anomalies in this condition. The aforementioned situation is reversed in NP8, when in the anomaly years local anomalies of the reverse type are favoured.

\begin{table}
	\centering
	\begin{tabular}{| c | c | c || c | c || c | c || c | c |}
		\hline
		Method      & NP  & NN   & STSP  & STSN  & SSP   & SSN   & TSP  & TSN \\
		\hline
		NP1	        & 185 & 76   &  9.4  & 13.9  &  4.6  &  4.6  & 2.5  & 2.9 \\
		NP2	        & 211 & 83   & 11.5  & 12.8  &  4.9  &  4.5  & 2.8  & 2.7    \\
		NP3	        & 185 & 111  &  9.9  & 12.9  &  4.6  &  4.4  & 2.6  & 3.4    \\
		NP4	        & 206 & 116  & 11.7  & 13.8  &  5.1  &  4.4  & 2.8  & 3.0   \\
		NP5	        & 186 &  75  & 10.1  & 14.1  &  4.6  &  4.7  & 2.7  & 2.9   \\
		NP6	        & 188 &  88  &  9.8  & 13.3  &  4.5  &  4.4  & 2.6  & 2.9   \\
		NP7	        & 189 &  80  & 11.1  & 12.8  &  5.1  &  4.6  & 2.7  & 2.8   \\
		NP8	        & 185 &  75  &  9.7  & 14.5  &  4.5  &  4.7  & 2.6  & 2.9   \\
		\hline
	\end{tabular}\caption{Mean spatial, temporal, and spatiotemporal sizes of positive and negative anomalies in different settings of node potentials of the MRF. The number of distinct positive anomalies is much larger than that of negative anomalies, while negative anomalies have larger mean spatiotemporal size, across these settings. Emphasizing node-specific potentials that depend on statistics at either the location or the year associated with the node (NP5-NP8) does not substantially alter these overall statistics.}
\end{table}

\end{document}